\documentclass[pdflatex,sn-apa]{sn-jnl}

\usepackage{graphicx}%
\usepackage{multirow}%
\usepackage{amsmath,amssymb,amsfonts}%
\usepackage{amsthm}%
\usepackage{mathrsfs}%
\usepackage[title]{appendix}%
\usepackage{xcolor}%
\usepackage{textcomp}%
\usepackage{manyfoot}%
\usepackage{booktabs}%
\usepackage{algorithm}%
\usepackage{algorithmicx}%
\usepackage{algpseudocode}%
\usepackage{listings}%
\usepackage{float}
\usepackage{placeins}
\usepackage{caption}

\theoremstyle{thmstyleone}%

\theoremstyle{thmstyletwo}%
\newtheorem{remark}{Remark}%

\theoremstyle{thmstylethree}%
\newtheorem{definition}{Definition}%

\begin{document}

\title[PWFs for Normal Tempered-Stable Laws]{Probability-Weighting Functions for Normal Tempered-Stable Laws: Orthogonal Channels and Information-Theoretic Indices}

\author[1]{\fnm{Akash} \sur{Deep}}\email{akash.deep@ttu.edu}

\author[1]{\fnm{Bhathiya} \sur{Divelgama}}\email{bdivelga@ttu.edu}

\author[1]{\fnm{A. Alexandre} \sur{Trindade}}\email{alex.trindade@ttu.edu}

\author[1]{\fnm{W. Brent} \sur{Lindquist}}\email{brent.lindquist@ttu.edu}

\author[1]{\fnm{Svetlozar T.} \sur{Rachev}}\email{zari.rachev@ttu.edu}

\author[2]{\fnm{Frank J.} \sur{Fabozzi}}\email{ffabozzi@jhu.edu}

\affil[1]{\orgdiv{Department of Mathematics and Statistics}, \orgname{Texas Tech University}, \orgaddress{\city{Lubbock}, \state{Texas}, \country{USA}}}

\affil[2]{\orgdiv{Carey Business School}, \orgname{Johns Hopkins University}, \orgaddress{\city{Baltimore}, \state{Maryland}, \country{USA}}}

\abstract{A probability-weighting function (PWF) maps objective probabilities into decision weights. We construct PWFs from changes in the distribution of returns rather than from a parametric family: with prior and posterior return distributions $F_{\text{prior}}$ and $F_{\text{post}}$, the PWF is $w(p) = F_{\text{post}}(F_{\text{prior}}^{-1}(p))$. Both distributions are normal tempered-stable (NTS), and the posterior differs from the prior along one of three channels, each isolated with the variance held fixed: scale, skewness and tail thickness. Around a symmetric benchmark with excess kurtosis 3, a 40\% change in scale moves cumulative probabilities by up to 7 to 8 percentage points, a skewness of $\pm 0.5$ by 2 percentage points, and a change of $\pm 0.3$ in the tail index by about 2 percentage points. The three channels leave distinct signatures: the deviation $w(p) - p$ changes sign once, twice and three times, respectively. Three information-theoretic indices (a Fisher-normalized logit shift, a signed Jensen--Shannon divergence and a log-odds elasticity) summarize the distortions and avoid the endpoint singularity of the classical ratio $(w(p)-p)/[p(1-p)]$. Applied to NTS parameters calibrated to S\&P 500 ETF daily returns over 2015--2025, the same channels produce distortions of 2 to 7 percentage points. Code that reproduces every figure and number in the paper is provided.}

\keywords{Probability-weighting function, Normal tempered-stable distribution, L\'evy subordination, Behavioral finance, Jensen--Shannon divergence, Greed--fear index}

\pacs[JEL Classification]{C46, C63, D81, G41}

\maketitle

\section*{Notation}
\begin{center}
\begin{tabular}{ll}
\hline
Symbol & Description \\
\hline
$T$ & Tempered-stable subordinator with $\mathbb{E}[T]=1$ \\
$Z$ & Standard normal variable independent of $T$ \\
$X$ & Normal tempered-stable (NTS) variable \\
$\alpha \in (0,2)$ & Tail index (Blumenthal--Getoor index of $X$) \\
$\theta > 0$ & Tempering parameter (smaller values give heavier tails) \\
$\beta$ & Asymmetry parameter \\
$\gamma > 0$ & Diffusive scale parameter \\
$\mu$ & Location parameter \\
$v = (2-\alpha)/(2\theta)$ & Variance of $T$ \\
$\ell_T(\lambda)$ & Laplace exponent of $T$ \\
$\phi_X(u)$ & Characteristic function of $X$ \\
$F_X,\ f_X,\ Q_X$ & Distribution function, density and quantile function of $X$ \\
$w(p)$ & Probability-weighting function \\
$\delta(p) = w(p) - p$ & PWF deviation \\
$g(p)$ & Classical greed--fear index $(w(p)-p)/[p(1-p)]$ \\
$G_{\text{FI}}(p)$ & Fisher-normalized logit-shift index \\
$S_{\text{JS}}(p)$ & Signed Jensen--Shannon index \\
$E(p)$ & Log-odds elasticity \\
$\mathcal{G},\ \mathcal{S},\ \mathcal{E}$ & Integrated indices \\
\hline
\end{tabular}
\end{center}

\section{Introduction}\label{sec:intro}

Probability-weighting functions (PWFs) describe how objective probabilities are transformed into the decision weights that enter a valuation. In behavioral finance they account for preferences for lotteries, insurance and tail protection that expected utility with linear weighting cannot explain \citep{kahneman1979prospect,tversky1992advances,prelec1998probability}, and in asset pricing they appear endogenously when probability weighting drives the demand for skewed securities \citep{barberis2008stocks}. In both settings, the curvature of the PWF, over and above the location, scale and tail features of the return distribution, is what produces differences in choices, prices and risk measures.

Most of the PWF literature postulates a parametric form, such as the one- or two-parameter families of \citet{tversky1992advances} and \citet{prelec1998probability}, and treats the return distribution as given. Return distributions, however, are neither Gaussian nor fixed: they show semi-heavy tails, skewness and jumps \citep{cont2001empirical}, and these features change with market conditions. L\'evy models obtained by subordinating Brownian motion to an independent increasing process reproduce the stylized facts while keeping the characteristic function in closed form \citep{barndorff1997processes,carr2002fine,cont2003financial}. The normal tempered-stable (NTS) family of \citet{kim2010tempered} and \citet{rachev2011financial} is a convenient member of this class: its five parameters control location, scale, asymmetry, tail index and tempering separately, and its moments are available in closed form.

This paper constructs PWFs from distributional perturbations instead of parametric assumptions. Given a prior (benchmark) return distribution $F_{\text{prior}}$ and a posterior $F_{\text{post}}$, the induced PWF is the quantile mapping $w(p) = F_{\text{post}}(Q_{\text{prior}}(p))$, where $Q_{\text{prior}} = F_{\text{prior}}^{-1}$. When both distributions are NTS, moving one parameter group at a time while holding the variance fixed isolates three channels, scale, skewness and tail thickness, and attaches a distributional cause to each PWF shape. The paper makes four contributions.

\textbf{1. Channel isolation.} We define three perturbation channels around any NTS benchmark, each holding the variance fixed, so that the resulting PWF reflects a single distributional feature. Because the quantile mapping is invariant to location and scale, the construction depends only on the shape parameters.

\textbf{2. Information-theoretic indices.} We introduce three indices of probability distortion, the Fisher-normalized logit shift $G_{\text{FI}}(p)$, the signed Jensen--Shannon divergence $S_{\text{JS}}(p)$ and the log-odds elasticity $E(p)$, and compare them with the classical greed--fear ratio $(w(p)-p)/[p(1-p)]$, whose endpoint singularity they avoid or control.

\textbf{3. Channel signatures.} Around a symmetric benchmark with excess kurtosis 3, the scale channel produces an inverse-S or S-shaped PWF with deviations of up to 7 to 8 percentage points, the skew channel a one-signed deviation of about 2 percentage points peaking at the median, and the tail channel a deviation of about 2 percentage points that changes sign three times. The number of sign changes of $w(p)-p$ therefore identifies the channel.

\textbf{4. Empirical calibration.} With NTS parameters fitted to daily returns on the SPDR S\&P 500 ETF over 2015--2025, the same channels produce distortions of 2 to 7 percentage points, and the fitted distribution reproduces the sample skewness and most of the excess kurtosis that a Gaussian cannot.

Our work draws on three literatures. The behavioral literature on probability weighting includes the parametric families of \citet{prelec1998probability} and \citet{tversky1992advances} and the information-theoretic measures of divergence between probability distributions \citep{csiszar1967information,lin1991divergence,endres2003new} that we adapt to the comparison of $w(p)$ with $p$. Closest to the present paper, \citet{deep2025probability} combine Student's $t$ return distributions with behavioral probability weighting in an econometric framework that preserves infinite divisibility; here the weighting function is instead derived from perturbations within the NTS family. The L\'evy-process literature provides the model: subordination \citep{barndorff1997processes,applebaum2009levy}, variance-gamma and normal inverse Gaussian laws \citep{madan1998variance,barndorff1997normal}, tempered-stable processes \citep{rosinski2007tempering}, and representations of tempered-stable processes as time-changed Brownian motion \citep{madan2008representing}. The asset-pricing literature documents that skewness and tail risk carry premia \citep{harvey1999autoregressive,bakshi2008stochastic}, which is one reason to ask how changes in skewness and tails translate into probability weights.

Section~\ref{sec:nts-background} sets out the NTS model, its moments and the numerical methods. Section~\ref{sec:pwf-theory} defines the PWF, the channels and the indices. Section~\ref{sec:cases} analyzes the three channels around a stylized benchmark. Section~\ref{sec:empirical} repeats the analysis around parameters calibrated to S\&P 500 returns. Section~\ref{sec:discussion} discusses implications, limitations and extensions. Numerical details are in the Appendix, and the code that generates every figure and number is available at the repository cited in the declarations.

\section{Normal Tempered-Stable Distributions}\label{sec:nts-background}

\subsection{Construction and Parameter Roles}\label{subsec:construction-roles}

The NTS distribution arises from Bochner's subordination principle: a Brownian motion observed at the random times given by an independent increasing L\'evy process (a subordinator) is again a L\'evy process, with tails and skewness inherited from the subordinator \citep{bochner1949diffusion,phillips1952generation}.

\begin{definition}[L\'evy Subordinator]
A L\'evy subordinator is a non-decreasing L\'evy process $T = (T_t)_{t \geq 0}$ with stationary, independent increments and c\`adl\`ag paths. Its law at time $t$ is determined by the Laplace exponent
\begin{equation}
\ell_T(\lambda) = d\lambda + \int_{(0,\infty)} (1 - e^{-\lambda x})\, \nu(dx), \qquad \mathbb{E}[e^{-\lambda T_t}] = e^{-t\,\ell_T(\lambda)},
\end{equation}
where $d \geq 0$ is the drift and $\nu$ is a L\'evy measure on $(0,\infty)$ with $\int_{(0,\infty)} (1 \wedge x)\,\nu(dx) < \infty$.
\end{definition}

Laplace exponents of subordinators are exactly the Bernstein functions \citep{schilling2012bernstein}. We use the tempered-stable subordinator in the parameterization of \citet{kim2010tempered}, normalized to unit mean so that the subordinator changes the shape but not the scale of the subordinated process.

\begin{definition}[Tempered-Stable Subordinator]\label{def:cts}
For $\alpha \in (0,2)$ and $\theta > 0$, the tempered-stable subordinator $T$ has L\'evy measure
\begin{equation}
\nu(dx) = \frac{\theta^{1-\alpha/2}}{\Gamma(1-\alpha/2)}\, x^{-1-\alpha/2}\, e^{-\theta x}\, \mathbf{1}_{x>0}\, dx
\end{equation}
and Laplace exponent
\begin{equation}
\ell_T(\lambda) = \frac{2\theta^{1-\alpha/2}}{\alpha}\left[(\theta+\lambda)^{\alpha/2} - \theta^{\alpha/2}\right].
\label{eq:laplace}
\end{equation}
Its index is $\alpha/2 \in (0,1)$, and the normalization gives $\mathbb{E}[T_1] = \ell_T'(0) = 1$ and
\begin{equation}
v := \operatorname{Var}(T_1) = -\ell_T''(0) = \frac{2-\alpha}{2\theta}.
\label{eq:varT}
\end{equation}
\end{definition}

The factor $e^{-\theta x}$ tempers the large jumps of the stable subordinator, so that all moments of $T$ are finite while the small-jump activity, governed by $\alpha/2$, is unchanged. Tempering is weaker for small $\theta$, which yields heavier tails and, by \eqref{eq:varT}, a more variable time change.

\begin{definition}[Normal Tempered-Stable Distribution]\label{def:nts}
Let $T$ be the subordinator of Definition~\ref{def:cts} at time one and let $Z \sim \mathcal{N}(0,1)$ be independent of $T$. For $\mu, \beta \in \mathbb{R}$ and $\gamma > 0$, the random variable
\begin{equation}
X = \mu + \beta\,(T - 1) + \gamma \sqrt{T}\, Z
\label{eq:nts}
\end{equation}
has the normal tempered-stable distribution, written $X \sim \text{NTS}(\alpha, \theta, \beta, \gamma, \mu)$. Its characteristic function is
\begin{equation}
\phi_X(u) = \mathbb{E}[e^{iuX}] = \exp\left\{ iu(\mu - \beta) - \frac{2\theta^{1-\alpha/2}}{\alpha}\left[\left(\theta - iu\beta + \tfrac{1}{2}u^2\gamma^2\right)^{\alpha/2} - \theta^{\alpha/2}\right]\right\}.
\label{eq:cf}
\end{equation}
\end{definition}

Equation \eqref{eq:cf} follows from \eqref{eq:laplace} by conditioning on $T$: given $T = t$, $X$ is Gaussian with mean $\mu + \beta(t-1)$ and variance $\gamma^2 t$, so $\phi_X(u) = e^{iu(\mu-\beta)} \exp\{-\ell_T(-iu\beta + u^2\gamma^2/2)\}$. The independence of $T$ and $Z$ is what makes $X$ a normal variance-mean mixture and keeps the characteristic function, the moments and the numerical inversion simple.

\begin{remark}[Special cases]
For $\alpha = 1$ the subordinator is inverse Gaussian and $X$ is normal inverse Gaussian \citep{barndorff1997normal}. As $\theta \to \infty$ the variance $v$ of $T$ tends to zero, $T \to 1$, and $X$ tends to $\mathcal{N}(\mu, \gamma^2)$; the Gaussian is therefore the limit of strong tempering. The Blumenthal--Getoor index of the L\'evy process generating $X$ is $\alpha$, twice the index of the subordinator.
\end{remark}

The five parameters have distinct roles. The location $\mu$ shifts the distribution without changing its shape. The scale $\gamma$ multiplies the Gaussian component and, when $\beta = 0$, is the standard deviation of $X$. The asymmetry parameter $\beta$ loads the random time on the mean: when $T$ is large, $X$ is pulled in the direction of $\beta$, so $\beta > 0$ produces a heavier right tail and positive skewness and $\beta < 0$ the opposite. The tail index $\alpha$ controls small-jump activity and, with $\theta$, the thickness of the tails: for fixed $\theta$, lower $\alpha$ increases $v$ and the excess kurtosis. The tempering parameter $\theta$ sets how quickly the tails decay; we hold it fixed throughout and use $\alpha$ as the tail channel.

\subsection{Moment Structure and Numerical Implementation}\label{subsec:moments-numerical}

Because $X$ is a variance-mean mixture, its cumulants follow from those of $T$. With $v$ as in \eqref{eq:varT}, the second, third and fourth cumulants of $T$ are $v$, $v(2-\alpha/2)/\theta$ and $v(2-\alpha/2)(3-\alpha/2)/\theta^2$, and
\begin{align}
\mathbb{E}[X] &= \mu, \label{eq:mean}\\
\operatorname{Var}(X) &= \gamma^2 + \beta^2 v, \label{eq:var}\\
\operatorname{Skew}(X) &= \frac{\beta^3\, v\,(2-\alpha/2)/\theta + 3\beta\gamma^2 v}{\operatorname{Var}(X)^{3/2}}, \label{eq:skew}\\
\operatorname{ExKurt}(X) &= \frac{\beta^4\, v\,(2-\alpha/2)(3-\alpha/2)/\theta^2 + 6\beta^2\gamma^2 v\,(2-\alpha/2)/\theta + 3\gamma^4 v}{\operatorname{Var}(X)^2}. \label{eq:kurt}
\end{align}
Two consequences organize the analysis in Sections~\ref{sec:cases} and~\ref{sec:empirical}. First, when $\beta = 0$ the variance equals $\gamma^2$ whatever the values of $\alpha$ and $\theta$, so the tail index can be changed at constant variance without any compensating adjustment, and the excess kurtosis is then $3v = 3(2-\alpha)/(2\theta)$. Second, for $\beta \neq 0$ the variance can be held at a target value $\sigma^2$ by setting $\gamma = (\sigma^2 - \beta^2 v)^{1/2}$, which requires $|\beta| < \sigma/\sqrt{v}$.

Densities, distribution functions and quantiles are computed from \eqref{eq:cf} by Fourier inversion, using the \texttt{temStaPy} library. The density is obtained from the inverse Fourier transform of $\phi_X$ evaluated by the fast Fourier transform on a uniform grid \citep{carr1999option}, and the distribution function from the Gil-Pelaez formula
\begin{equation}
F_X(x) = \frac{1}{2} - \frac{1}{\pi} \int_0^\infty \operatorname{Im}\!\left[\frac{e^{-iux}\phi_X(u)}{u}\right] du,
\label{eq:gilpelaez}
\end{equation}
also evaluated by FFT \citep{gilpelaez1951note}. Both are interpolated between grid points by monotone piecewise-cubic (PCHIP) interpolation, and the quantile function is the PCHIP inverse of the tabulated distribution function. Grid sizes and ranges are listed in Appendix~\ref{app:numerics}.

\subsection{Visualizing NTS Parameter Effects}\label{subsec:visualization}

\begin{figure}[!tbp]
\centering
\includegraphics[width=0.7\textwidth]{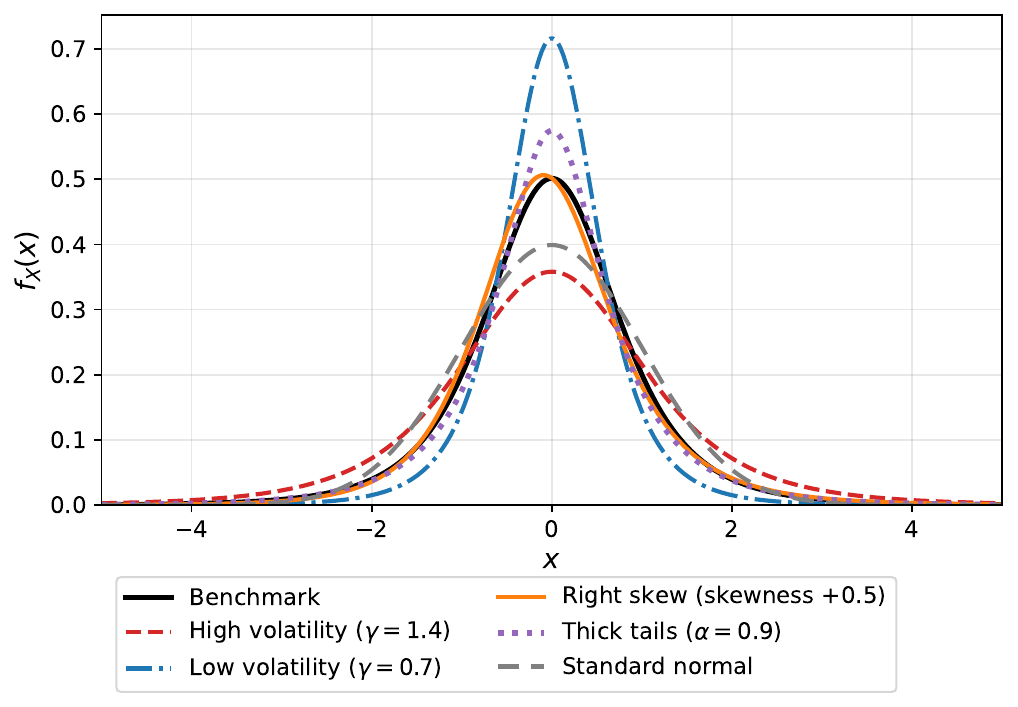}
\caption{Densities of the benchmark NTS \eqref{eq:benchmark}, four single-channel perturbations and the standard normal.}
\label{fig:pdfs}
\end{figure}

\begin{figure}[!tbp]
\centering
\includegraphics[width=0.7\textwidth]{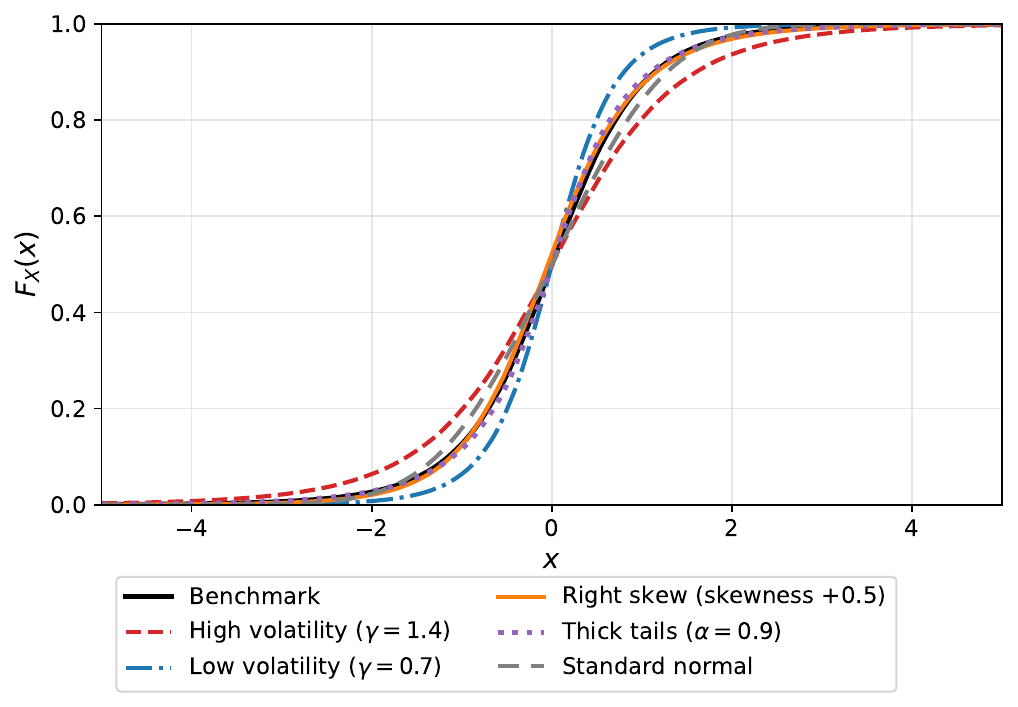}
\caption{Distribution functions corresponding to the densities of Fig.~\ref{fig:pdfs}, with the same colour scheme.}
\label{fig:cdfs}
\end{figure}

\begin{figure}[!tbp]
\centering
\includegraphics[width=0.7\textwidth]{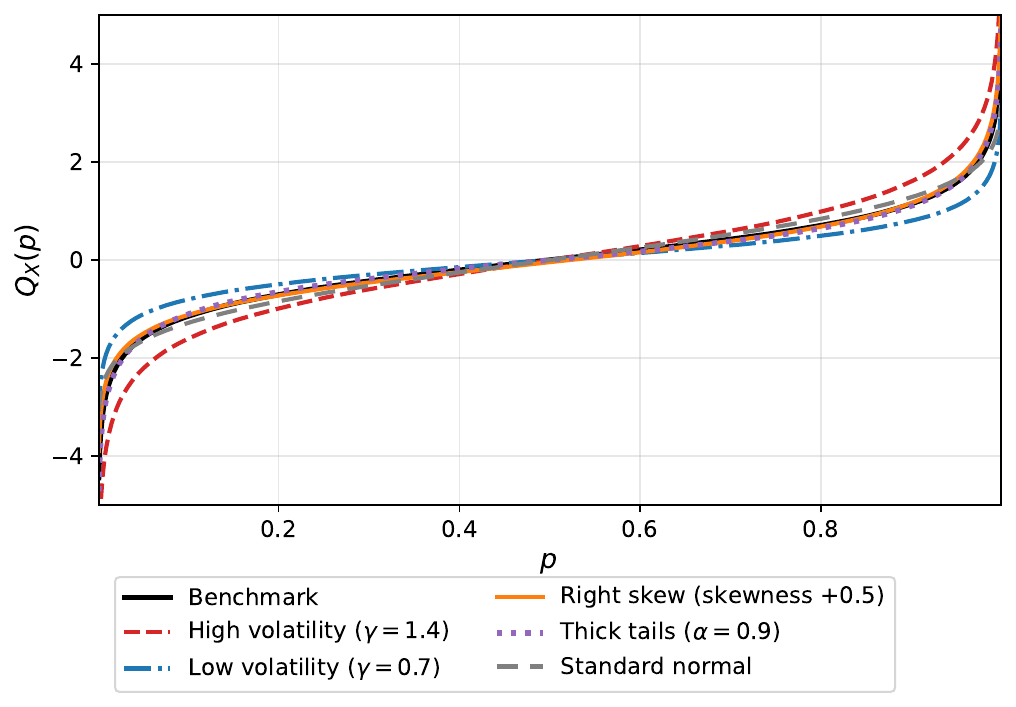}
\caption{Quantile functions $Q_X(p)$ corresponding to the densities of Fig.~\ref{fig:pdfs}, with the same colour scheme.}
\label{fig:quantiles}
\end{figure}

Figures~\ref{fig:pdfs}--\ref{fig:quantiles} show how each channel changes the distribution, starting from the benchmark used in Section~\ref{sec:cases},
\begin{equation}
\text{benchmark: } \text{NTS}(\alpha = 1.2,\ \theta = 0.4,\ \beta = 0,\ \gamma = 1,\ \mu = 0),
\label{eq:benchmark}
\end{equation}
which has unit variance, zero skewness and excess kurtosis 3 by \eqref{eq:var}--\eqref{eq:kurt}. Fig.~\ref{fig:pdfs} plots its density (black solid) together with a high-volatility variant ($\gamma = 1.4$, red dashed), a low-volatility variant ($\gamma = 0.7$, blue dash-dotted), a right-skewed variant with skewness $+0.5$ at unit variance (orange solid), a thick-tailed variant ($\alpha = 0.9$, purple dotted) and the standard normal density (gray dashed). Relative to the normal, the benchmark has a higher peak and heavier tails, the signature of excess kurtosis. Changing $\gamma$ rescales the density; the skewed variant tilts it to the right; the thick-tailed variant sharpens the peak and raises both tails while leaving the shoulders lower, as a change in kurtosis at fixed variance must.

Fig.~\ref{fig:cdfs} shows the corresponding distribution functions with the same colour scheme. Volatility changes alter the slope at the median, the skewed variant moves the median to the left of zero and gains mass in the upper tail, and the thick-tailed variant crosses the benchmark three times, once near the median and once in each tail.

Fig.~\ref{fig:quantiles} plots the quantile functions $Q_X(p) = F_X^{-1}(p)$, which are the ingredients of the PWF construction $w(p) = F_{\text{post}}(Q_{\text{prior}}(p))$ of Section~\ref{sec:pwf-theory}. High volatility spreads the quantiles apart and low volatility compresses them; the skewed variant lowers the central quantiles and raises the upper ones; the thick-tailed variant lies inside the benchmark in the shoulders and outside it in the extremes. These deformations are what produce PWF curvature.

\FloatBarrier

\section{Probability-Weighting Functions and Information-Theoretic Indices}\label{sec:pwf-theory}

\subsection{Construction and Channel Isolation}\label{subsec:construction-channels}

We obtain probability-weighting functions from a change of distribution rather than from a parametric family. The PWF then has a cause: it is the distortion that a specific change in beliefs about returns imposes on cumulative probabilities.

\begin{definition}[Quantile-Based PWF]\label{def:pwf}
Let $F_{\text{prior}}$ and $F_{\text{post}}$ be continuous, strictly increasing distribution functions and let $Q_{\text{prior}} = F_{\text{prior}}^{-1}$. The probability-weighting function induced by the transition from prior to posterior is
\begin{equation}
w(p) = F_{\text{post}}\!\left( Q_{\text{prior}}(p) \right), \qquad p\in[0,1].
\label{eq:pwf}
\end{equation}
\end{definition}

The function $w$ is increasing with $w(0) = 0$ and $w(1) = 1$, and $w(p) = p$ for all $p$ if and only if $F_{\text{prior}} = F_{\text{post}}$. Definition~\ref{def:pwf} is a single construction, but it generates a family of weighting functions, one for each posterior; the paper studies six of them in Section~\ref{sec:cases} and six more in Section~\ref{sec:empirical}. In this sense the definition is generic, as the definition of a distribution function is generic: it admits many functions, and the question is which of them are plausible. The pairs studied here are those in which prior and posterior are both NTS and hence infinitely divisible, so that the same laws can serve as return distributions in asset pricing; this is the case of interest for connecting probability weighting with rational finance. The derivative of $w$ is the ratio of the two densities at the prior quantile,
\begin{equation}
w'(p) = \frac{f_{\text{post}}(Q_{\text{prior}}(p))}{f_{\text{prior}}(Q_{\text{prior}}(p))},
\label{eq:wprime}
\end{equation}
which we use to compute the elasticity index below without numerical differentiation.

\begin{remark}[Invariance]
If $h$ is strictly increasing, the PWF for the pair $(h(X_{\text{prior}}), h(X_{\text{post}}))$ equals the PWF for $(X_{\text{prior}}, X_{\text{post}})$. In particular, $w$ is unchanged when both distributions are shifted by the same amount or multiplied by the same positive constant, so only the shape of the posterior relative to the prior matters. For NTS pairs this means that $w$ depends on $(\alpha, \theta)$, on the ratio $\gamma_{\text{post}}/\gamma_{\text{prior}}$, and on $\beta/\gamma$, not on the absolute scale or location.
\end{remark}

To attribute PWF shapes to distributional features we change one parameter group at a time and keep the variance fixed, so that a change in dispersion never masquerades as a change in asymmetry or in tail weight.

\begin{definition}[Orthogonal Parameter Channels]\label{def:channels}
Given a benchmark $\text{NTS}(\alpha, \theta, \beta, \gamma, \mu)$ with variance $\sigma^2$ and skewness $s$, the three channels are as follows. The \textit{scale channel} multiplies $\gamma$ by a factor $c > 0$ with $(\alpha, \theta, \beta, \mu)$ fixed. The \textit{skew channel} changes the skewness to $s \pm \Delta$ by choosing $\beta$ from \eqref{eq:skew}, with $(\alpha, \theta, \mu)$ fixed and $\gamma$ reset so that the variance \eqref{eq:var} remains $\sigma^2$. The \textit{tail channel} changes $\alpha$ by $\pm\Delta_\alpha$ with $(\theta, \beta, \mu)$ fixed and $\gamma$ reset to keep the variance at $\sigma^2$; when $\beta = 0$ no reset is needed.
\end{definition}

For each channel we call the direction that worsens the outlook \textit{fearful} (higher dispersion, more negative skewness, heavier tails) and the opposite direction \textit{greedy}. The labels are mnemonic: they describe the posterior belief, not a utility function.

\subsection{Information-Theoretic Indices}\label{subsec:info-indices}

The classical greed--fear index
\begin{equation}
g(p) = \frac{w(p) - p}{p(1-p)}, \quad p \in (0,1),
\label{eq:classical}
\end{equation}
normalizes the deviation by $p(1-p)$ so that distortions near the endpoints, where $w(p) - p$ must vanish, are not mechanically suppressed. The normalization has a cost: whenever the tails of the posterior differ from those of the prior, $w(p) - p$ vanishes more slowly than $p(1-p)$ as $p \to 0$ or $1$, and $g(p)$ diverges. We therefore complement $g$ with three indices built from the log-odds and from an information divergence.

The first index works in log-odds space, where additive shifts have a multiplicative interpretation. With $\operatorname{logit}(p) = \log[p/(1-p)]$, the pointwise log-odds gap between subjective and objective probabilities is
\begin{equation}
G(p) = \operatorname{logit}(w(p)) - \operatorname{logit}(p) = \log\frac{w(p)/(1-w(p))}{p/(1-p)},
\end{equation}
and its Fisher-normalized version is
\begin{equation}
G_{\text{FI}}(p) = \frac{G(p)}{\sqrt{p(1-p)}},
\label{eq:gfi}
\end{equation}
where $p(1-p)$ is the reciprocal of the Fisher information of a Bernoulli trial with success probability $p$, so that $G_{\text{FI}}$ measures the log-odds gap in units of the standard error of a single observation. Positive values mean that the posterior overweights the event $\{X \leq Q_{\text{prior}}(p)\}$, negative values that it underweights it. Like $g$, $G_{\text{FI}}$ is unbounded at the endpoints when the tails differ, but it grows only like $|G(p)|/\sqrt{p}$ rather than like $|w(p)-p|/p$, and the figures below cut its axis accordingly.

The second index is bounded. For Bernoulli distributions with parameters $w(p)$ and $p$, the Jensen--Shannon divergence \citep{lin1991divergence} is
\begin{equation}
\text{JSD}(p) = \tfrac{1}{2}\,\text{KL}\!\left(w(p) \parallel m\right) + \tfrac{1}{2}\,\text{KL}\!\left(p \parallel m\right), \qquad m = \tfrac{1}{2}\,[w(p) + p],
\end{equation}
where $\text{KL}(a \parallel b) = a\log(a/b) + (1-a)\log[(1-a)/(1-b)]$ with natural logarithms. Its square root is a metric \citep{endres2003new}, and we attach the sign of the deviation:
\begin{equation}
S_{\text{JS}}(p) = \operatorname{sgn}\big(w(p) - p\big)\, \sqrt{2\,\text{JSD}(p)}.
\label{eq:sjs}
\end{equation}
Since $\text{JSD} \leq \log 2$, $S_{\text{JS}}(p) \in [-\sqrt{2\log 2}, \sqrt{2\log 2}] \approx [-1.18, 1.18]$, with the bounds attained only when one probability is 0 and the other 1. The index is smooth, increases with $|w(p) - p|$ and vanishes at the endpoints, so it is suited to comparisons across channels and across assets.

The third index measures where the distortion is accelerating. The log-odds elasticity is the derivative of the log-odds gap,
\begin{equation}
E(p) = \frac{d}{dp}\bigl[\operatorname{logit}(w(p)) - \operatorname{logit}(p)\bigr] = \frac{w'(p)}{w(p)(1-w(p))} - \frac{1}{p(1-p)},
\label{eq:elasticity}
\end{equation}
with $w'(p)$ from \eqref{eq:wprime}. Positive values mean that subjective odds rise faster than objective odds as $p$ increases; $E$ changes sign at the inflection points of $G$. In all computations the probabilities entering the logarithms are clipped to $[\varepsilon, 1-\varepsilon]$ with $\varepsilon = 10^{-6}$.

For scalar summaries we integrate over $p$:
\begin{equation}
\mathcal{G} = \int_0^1 G_{\text{FI}}(p)\,dp, \quad
\mathcal{S} = \int_0^1 S_{\text{JS}}(p)\,dp, \quad
\mathcal{E} = \int_0^1 \lvert E(p)\rvert\,dp,
\label{eq:aggregates}
\end{equation}
and also report the absolute versions $|\mathcal{G}|_1 = \int_0^1 |G_{\text{FI}}(p)|\,dp$ and $|\mathcal{S}|_1 = \int_0^1 |S_{\text{JS}}(p)|\,dp$. The signed aggregates measure net overweighting of low outcomes against high ones; when prior and posterior are both symmetric about the same point, $w(1-p) = 1 - w(p)$, the pointwise indices are antisymmetric about $p = 1/2$, and $\mathcal{G}$ and $\mathcal{S}$ vanish identically. The absolute aggregates and $\mathcal{E}$ measure the size of the distortion regardless of direction.

\FloatBarrier

\section{Computational Analysis: Three Orthogonal Channels}\label{sec:cases}

We apply Definition~\ref{def:channels} to the benchmark \eqref{eq:benchmark} with scale factors $c \in \{1.4, 0.7\}$, a skewness shift $\Delta = 0.5$ and a tail shift $\Delta_\alpha = 0.3$. Table~\ref{tab:stylized_specs} lists the seven specifications with their implied moments. Within each channel, the fearful and greedy specifications have the same variance as the benchmark except, by construction, in the scale channel. All PWFs are evaluated on the grid $p_i = i/2001$, $i = 1, \dots, 2000$.

\begin{table}[htbp]
\centering
\caption{Benchmark and channel specifications of Section~\ref{sec:cases} with implied moments}
\label{tab:stylized_specs}
\begin{tabular}{llccccccc}
\toprule
Channel & Specification & $\alpha$ & $\theta$ & $\beta$ & $\gamma$ & Std.\ dev. & Skewness & Excess kurtosis \\
\midrule
--    & Benchmark & 1.2 & 0.4 & 0 & 1 & 1.00 & 0 & 3.00 \\
Scale & Fearful   & 1.2 & 0.4 & 0 & 1.4 & 1.40 & 0 & 3.00 \\
Scale & Greedy    & 1.2 & 0.4 & 0 & 0.7 & 0.70 & 0 & 3.00 \\
Skew  & Fearful   & 1.2 & 0.4 & $-0.166$ & 0.986 & 1.00 & $-0.50$ & 3.42 \\
Skew  & Greedy    & 1.2 & 0.4 & $+0.166$ & 0.986 & 1.00 & $+0.50$ & 3.42 \\
Tail  & Fearful   & 0.9 & 0.4 & 0 & 1 & 1.00 & 0 & 4.13 \\
Tail  & Greedy    & 1.5 & 0.4 & 0 & 1 & 1.00 & 0 & 1.88 \\
\bottomrule
\end{tabular}
\end{table}

\medskip
\noindent\textbf{Case 1: Scale Channel}\label{subsec:case1}

The scale channel multiplies $\gamma$ by 1.4 (fearful) or 0.7 (greedy) and leaves $\alpha$, $\theta$ and $\beta$ unchanged, so the standard deviation becomes 1.4 or 0.7 while skewness and excess kurtosis are unchanged (Table~\ref{tab:stylized_specs}). By the invariance remark, the resulting PWF depends on the scale ratio only.

\begin{figure}[!tbp]
\centering
\includegraphics[width=0.7\textwidth]{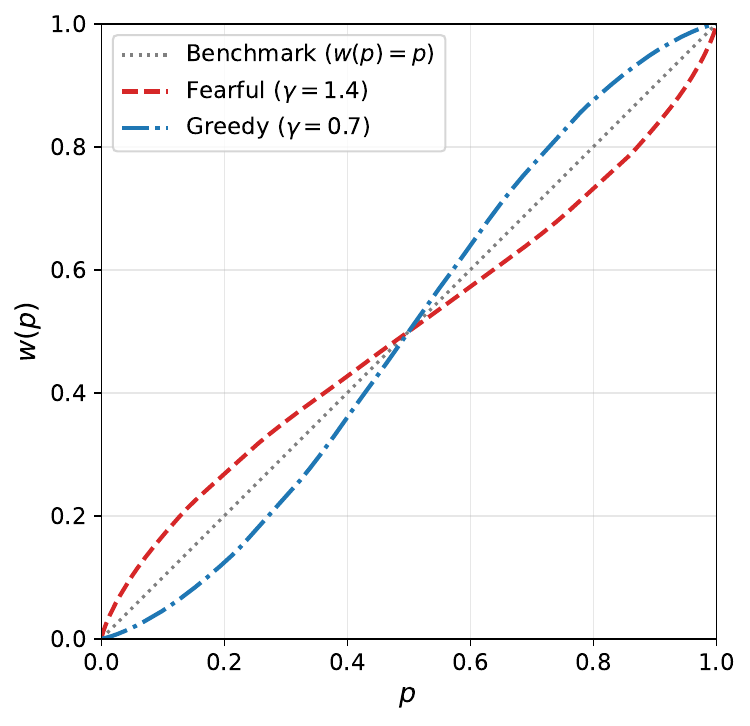}
\caption{Probability-weighting functions for the two specifications of the scale channel.}
\label{fig:case1_pwf}
\end{figure}

Fig.~\ref{fig:case1_pwf} plots the two PWFs. The high-dispersion (fearful) curve has the inverse-S shape of prospect theory: a posterior with more spread puts more mass below any negative prior quantile and less above any positive one, so it overweights small cumulative probabilities and underweights large ones. The largest deviation is $7.3$ percentage points at $p = 0.14$. The low-dispersion (greedy) curve is S-shaped, with a largest deviation of $7.7$ percentage points at $p = 0.22$. Both curves cross the diagonal exactly once, at $p = 1/2$, because prior and posterior are symmetric about the same point.

\begin{figure}[!tbp]
\centering
\includegraphics[width=0.7\textwidth]{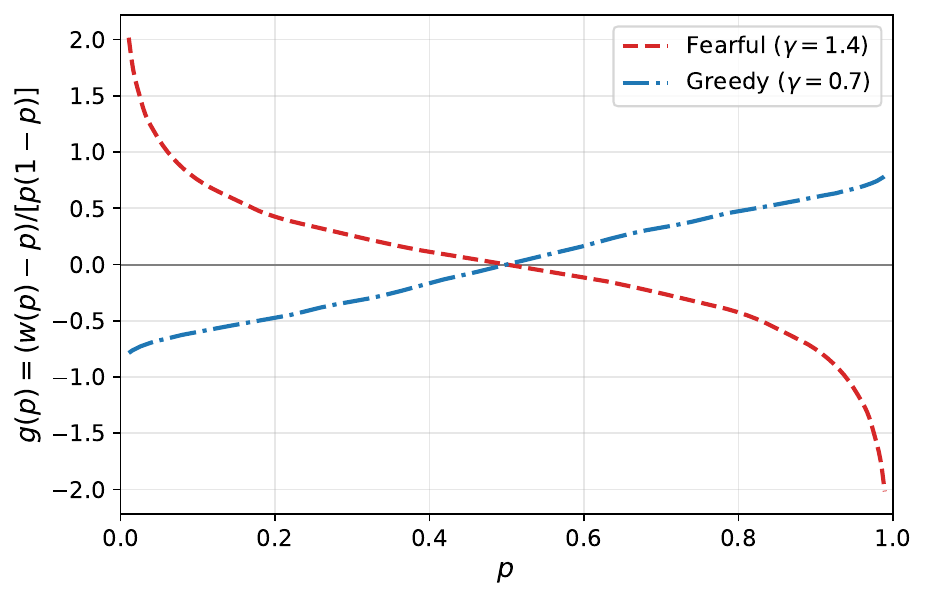}
\caption{Classical greed--fear index $g(p)$ for the two specifications of the scale channel, on $p \in [0.01, 0.99]$.}
\label{fig:case1_classical}
\end{figure}

Fig.~\ref{fig:case1_classical} plots the classical index $g(p)$ on $[0.01, 0.99]$. The fearful curve is positive below the median and negative above it, with $g(0.05) = 1.1$; the greedy curve has the opposite sign pattern, with $g(0.05) = -0.7$. Near the endpoints the fearful curve grows without bound, reaching $\pm 2.0$ at $p = 0.01$ and $0.99$ and continuing to rise outside the plotted range, which is the singularity that motivates the indices of Section~\ref{subsec:info-indices}.

\begin{figure}[!tbp]
\centering
\includegraphics[width=0.7\textwidth]{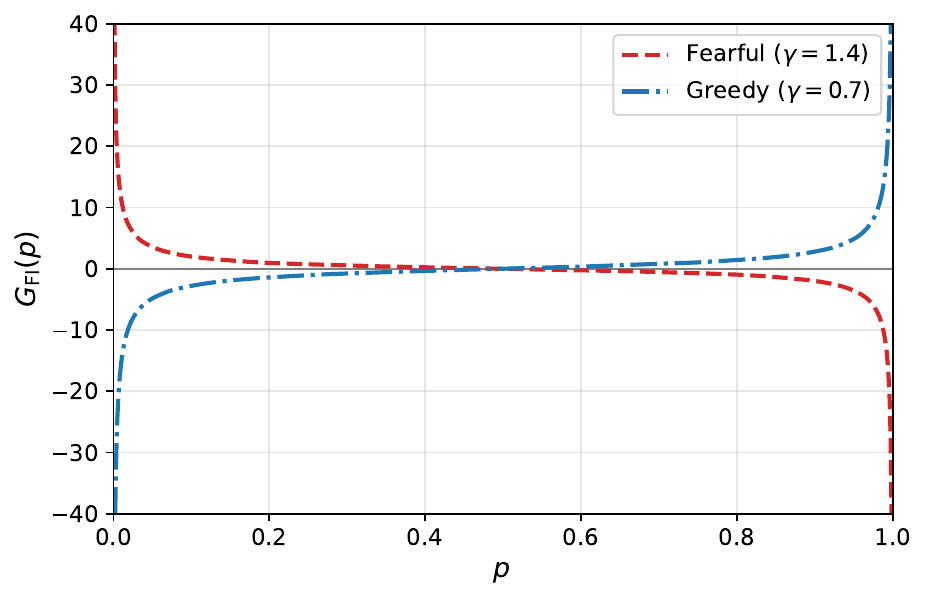}
\caption{Fisher-normalized logit-shift index $G_{\text{FI}}(p)$ for the two specifications of the scale channel; the vertical axis is cut at $\pm 40$.}
\label{fig:case1_logit}
\end{figure}

Fig.~\ref{fig:case1_logit} plots the Fisher-normalized logit shift $G_{\text{FI}}(p)$ on the full grid, with the vertical axis cut at $\pm 40$. The index is close to zero over the central range, where the log-odds gap is small, and grows towards the endpoints: $G_{\text{FI}}(0.01) = 11.3$ for the fearful curve and $-15.4$ for the greedy curve, with the signs reversed at $p = 0.99$. The sign pattern is the same as that of $g$, but the normalization by $\sqrt{p(1-p)}$ rather than $p(1-p)$ keeps the index finite over a wider range and makes the two tails comparable in units of one observation's standard error. The signed aggregate $\mathcal{G}$ is zero for both specifications by antisymmetry; the absolute aggregates are $|\mathcal{G}|_1 = 1.68$ (fearful) and $2.34$ (greedy), $|\mathcal{S}|_1 = 0.062$ and $0.073$, and $\mathcal{E} = 3.59$ and $4.98$. The bounded index $S_{\text{JS}}$ reaches at most $0.10$ and $0.11$ in absolute value.

\FloatBarrier

\medskip
\noindent\textbf{Case 2: Skew Channel}\label{subsec:case2}

The skew channel sets the posterior skewness to $-0.5$ (fearful) or $+0.5$ (greedy) by solving \eqref{eq:skew} for $\beta$, which gives $\beta = \mp 0.166$, and resets $\gamma$ to $0.986$ so that the variance stays at 1 (Table~\ref{tab:stylized_specs}). The excess kurtosis rises slightly, from 3.00 to 3.42, because the term $\beta^2 v$ in \eqref{eq:var} replaces part of the Gaussian variance by subordinator variance; this is the price of holding the variance fixed while changing $\beta$, and it is small.

\begin{figure}[!tbp]
\centering
\includegraphics[width=0.7\textwidth]{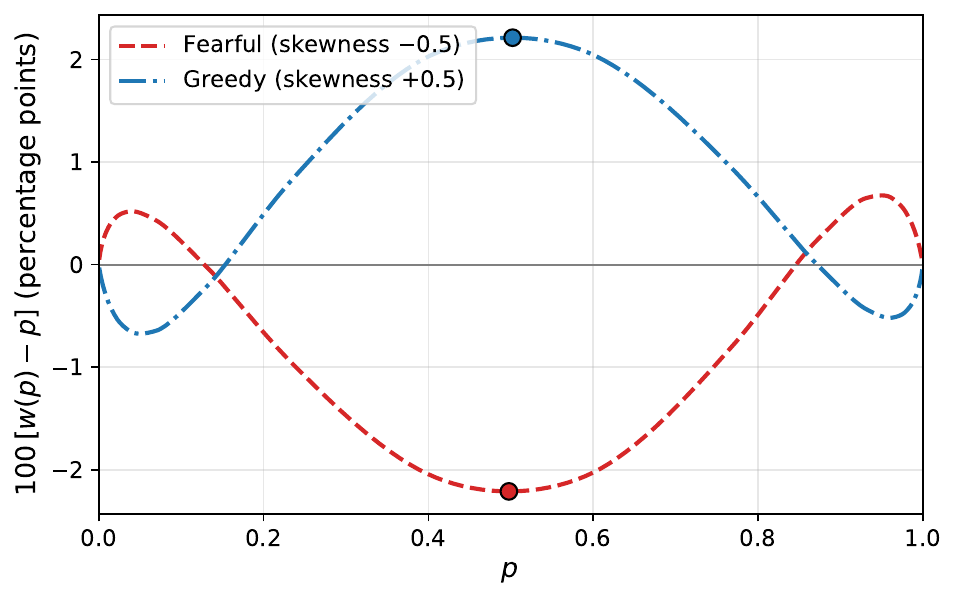}
\caption{PWF deviation $\delta(p) = w(p) - p$ for the two specifications of the skew channel; markers show the largest deviations.}
\label{fig:case2_deviation}
\end{figure}

Fig.~\ref{fig:case2_deviation} plots the deviation $\delta(p) = w(p) - p$ in percentage points. The two curves are mirror images, $\delta_{\text{greedy}}(p) = -\delta_{\text{fearful}}(1-p)$, and differ in kind from the scale channel: the deviation is one-signed over most of the range and peaks at the median, where it reaches $-2.2$ percentage points for the fearful specification and $+2.2$ for the greedy one. A left-skewed posterior has its median above the mean, so it assigns less than $1/2$ to the event $\{X \leq 0\}$ and, more generally, less probability than the benchmark to every central prior quantile; the right-skewed posterior does the opposite. Close to the endpoints the sign reverses: the left-skewed posterior has a heavier left tail and a lighter right tail than the symmetric benchmark, so it overweights both $\{X \leq Q_{\text{prior}}(p)\}$ for $p$ near 0 and its complement for $p$ near 1, and $\delta$ crosses zero twice, at $p \approx 0.14$ and $0.86$. The reversal is small, below $0.7$ percentage points, but it dominates the endpoint-weighted index $G_{\text{FI}}$, which is positive at both ends for the fearful specification ($G_{\text{FI}}(0.01) = 2.9$, $G_{\text{FI}}(0.99) = 4.1$). The signed aggregates are therefore not zero here, $\mathcal{G} = +0.21$ and $\mathcal{S} = -0.008$ for the fearful specification and the negatives for the greedy one, and they carry opposite signs: $\mathcal{S}$, which weights the centre, records the central underweighting, while $\mathcal{G}$, which weights the tails, records the tail reversal.

\FloatBarrier

\medskip
\noindent\textbf{Case 3: Tail-Thickness Channel}\label{subsec:case3}

The tail channel lowers $\alpha$ to $0.9$ (fearful, excess kurtosis 4.13) or raises it to $1.5$ (greedy, excess kurtosis 1.88) with $\theta$, $\beta$ and $\gamma$ unchanged; because $\beta = 0$, the variance is exactly 1 in both cases by \eqref{eq:var}. The channel therefore changes kurtosis and nothing else.

\begin{figure}[!tbp]
\centering
\includegraphics[width=0.7\textwidth]{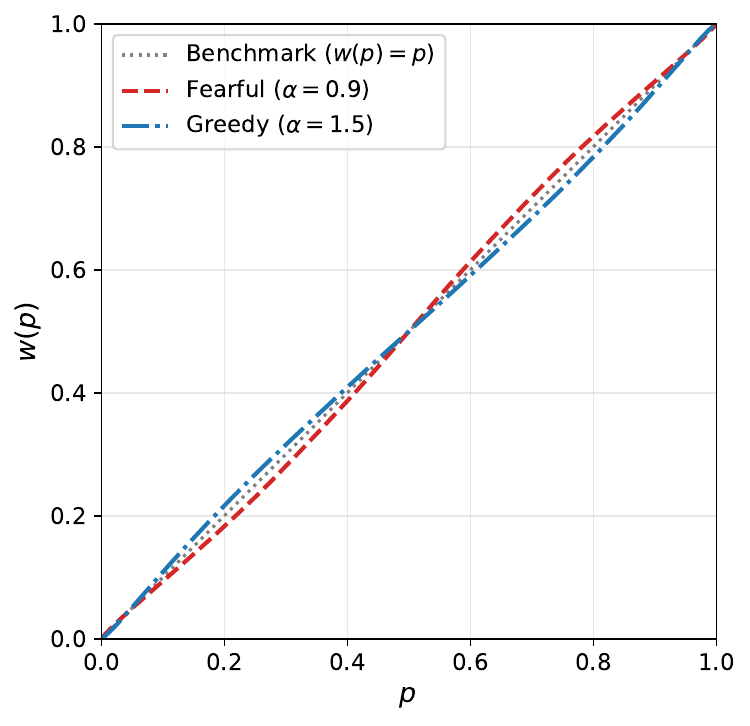}
\caption{Probability-weighting functions for the two specifications of the tail-thickness channel.}
\label{fig:case3_pwf}
\end{figure}

Fig.~\ref{fig:case3_pwf} plots the two PWFs. Both stay close to the diagonal, with largest deviations of $2.0$ percentage points at $p = 0.26$ (fearful) and $1.7$ percentage points at $p = 0.76$ (greedy), and both cross it three times, at $p \approx 0.1$, $0.5$ and $0.9$. A heavier-tailed posterior with the same variance has more mass in the centre and in the far tails and less in the shoulders: relative to the benchmark it overweights the extreme events, underweights the event $\{X \leq Q_{\text{prior}}(p)\}$ for $p$ in the lower shoulder, and overweights it in the upper shoulder. The thin-tailed posterior reverses each sign.

\begin{figure}[!tbp]
\centering
\includegraphics[width=0.7\textwidth]{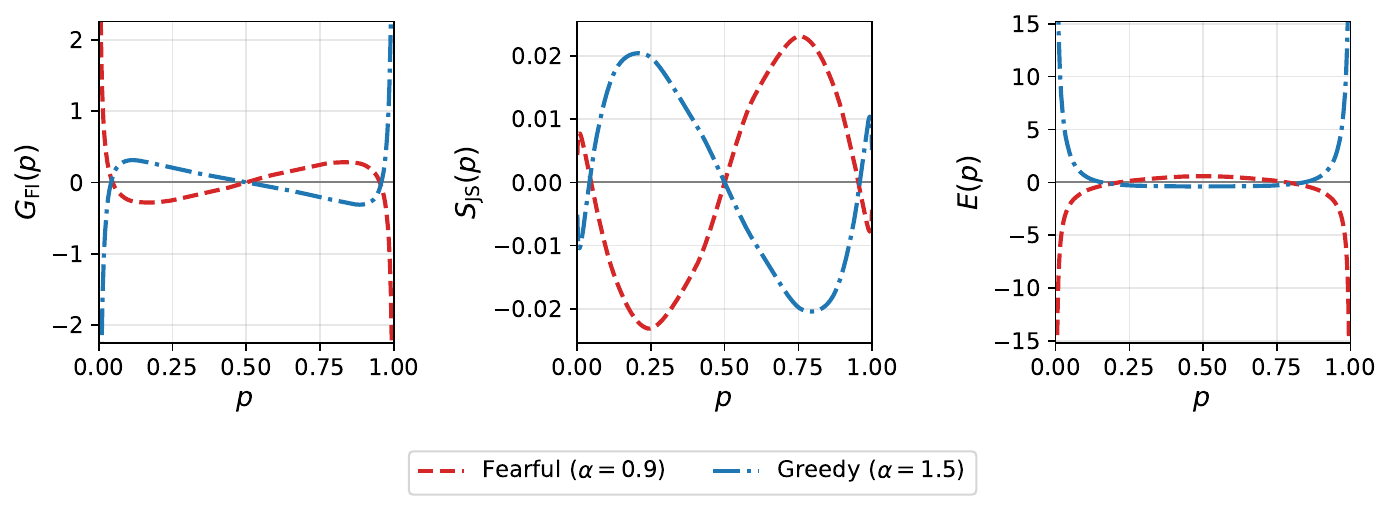}
\caption{Indices $G_{\text{FI}}$, $S_{\text{JS}}$ and $E$ for the two specifications of the tail-thickness channel; the outer panels are cut at the range attained on $[0.01, 0.99]$.}
\label{fig:case3_info}
\end{figure}

Fig.~\ref{fig:case3_info} plots the three indices for the same two specifications. In the interior all three are small, with $|G_{\text{FI}}| \leq 0.5$, $|S_{\text{JS}}| \leq 0.023$ and $|E| \leq 2.5$ on $[0.05, 0.95]$, and each changes sign three times, following $\delta$. Towards the endpoints $G_{\text{FI}}$ and $E$ grow, since the tails of the two distributions differ by construction: $G_{\text{FI}}(0.01) = 1.5$ for the fearful and $-2.2$ for the greedy specification. The absolute aggregates, $|\mathcal{G}|_1 = 0.27$ and $0.31$, $|\mathcal{S}|_1 = 0.014$ and $0.013$, and $\mathcal{E} = 1.16$ and $1.47$, are between a sixth and a third of their scale-channel counterparts.

\medskip
\noindent\textbf{Channel signatures}

The three channels can be told apart from the deviation $\delta(p)$ alone. A change in scale produces one sign change, at the median; a change in skewness at fixed variance produces two, with a one-signed body and small reversals in the tails; a change in tail thickness at fixed variance produces three. The magnitudes also differ: for the perturbations in Table~\ref{tab:stylized_specs}, which are of comparable economic size, the scale channel moves cumulative probabilities by up to 7 to 8 percentage points, the skew channel by 2, and the tail channel by about 2. Section~\ref{sec:empirical} shows that the same ordering and the same sign patterns hold around parameters estimated from market data.

\FloatBarrier

\section{Empirical Application: S\&P 500 Returns}\label{sec:empirical}

We now calibrate the benchmark to equity returns and apply the three channels around the estimated parameters. The exercise serves two purposes: it checks that the NTS family describes the data well enough for the channels to be meaningful, and it gives the PWF distortions of Section~\ref{sec:cases} an empirical scale.

\subsection{Data and Estimation}\label{subsec:data}

We use daily log-returns $r_t = \ln(P_t/P_{t-1})$ on the SPDR S\&P 500 ETF Trust (SPY), where $P_t$ is the adjusted closing price from Bloomberg, from 2 January 2015 to 3 December 2025, giving $N = 2{,}746$ observations. The ETF is a liquid, tradeable proxy for the index with small tracking error. Table~\ref{tab:sample_stats} reports the sample statistics: the mean is close to zero, the annualized volatility is 17.9\%, the skewness is $-0.58$ and the excess kurtosis is $14.5$. The minimum of $-11.6\%$ and the maximum of $+10.0\%$ both date from March 2020.

We estimate the five NTS parameters with the \texttt{fitnts} routine of \texttt{temStaPy}. The routine sets $\mu$ and the standard deviation to the sample mean and standard deviation, standardizes the data, and chooses the shape parameters $(\alpha, \theta, \beta/\sigma)$ to minimize the mean squared distance between the NTS distribution function and a kernel-smoothed empirical distribution function on the standardized sample; $\gamma$ then follows from \eqref{eq:var}. This is a minimum-distance estimator in the distribution-function domain, not maximum likelihood; it is fast, stable in the presence of the extreme observations in the sample, and adequate for a calibration whose purpose is to anchor the channel analysis. Table~\ref{tab:nts_params} reports the estimates and the moments they imply.

\begin{table}[htbp]
\centering
\begin{minipage}[t]{0.46\textwidth}
\centering
\captionof{table}{Sample statistics, SPY daily log-returns, 2015--2025}
\label{tab:sample_stats}
\begin{tabular}{lc}
\toprule
Statistic & Value \\
\midrule
Observations & 2,746 \\
Mean & 0.000505 \\
Standard deviation & 0.01126 \\
Annualized volatility & 17.9\% \\
Skewness & $-0.583$ \\
Excess kurtosis & 14.48 \\
Minimum & $-0.1159$ \\
Maximum & $+0.0999$ \\
\bottomrule
\end{tabular}
\end{minipage}\hfill
\begin{minipage}[t]{0.50\textwidth}
\centering
\captionof{table}{NTS estimates and implied moments}
\label{tab:nts_params}
\begin{tabular}{llc}
\toprule
Parameter & Symbol & Estimate \\
\midrule
Tail index & $\alpha$ & 1.070 \\
Tempering & $\theta$ & 0.204 \\
Asymmetry & $\beta$ & $-0.00088$ \\
Scale & $\gamma$ & 0.01118 \\
Location & $\mu$ & 0.000505 \\
\midrule
Implied std.\ dev. & & 0.01126 \\
Implied skewness & & $-0.53$ \\
Implied excess kurtosis & & 7.23 \\
\bottomrule
\end{tabular}
\end{minipage}
\end{table}

The estimated tail index $\hat\alpha = 1.07$ is close to the normal inverse Gaussian case $\alpha = 1$ and within the range typically reported for equity returns \citep{carr2002fine}. The tempering parameter $\hat\theta = 0.20$ is small, so the subordinator is highly variable ($v = 2.28$) and the fitted distribution is strongly leptokurtic. The asymmetry parameter is small in absolute terms, $\hat\beta = -0.00088$, but relative to the scale it is not: $\hat\beta/\hat\gamma = -0.08$, which with $v = 2.28$ produces an implied skewness of $-0.53$, close to the sample value. The implied excess kurtosis of $7.2$ is about half the sample value of $14.5$; the sample value is driven by a handful of returns in March 2020 and is itself imprecisely estimated.

\begin{figure}[!tbp]
\centering
\includegraphics[width=0.7\textwidth]{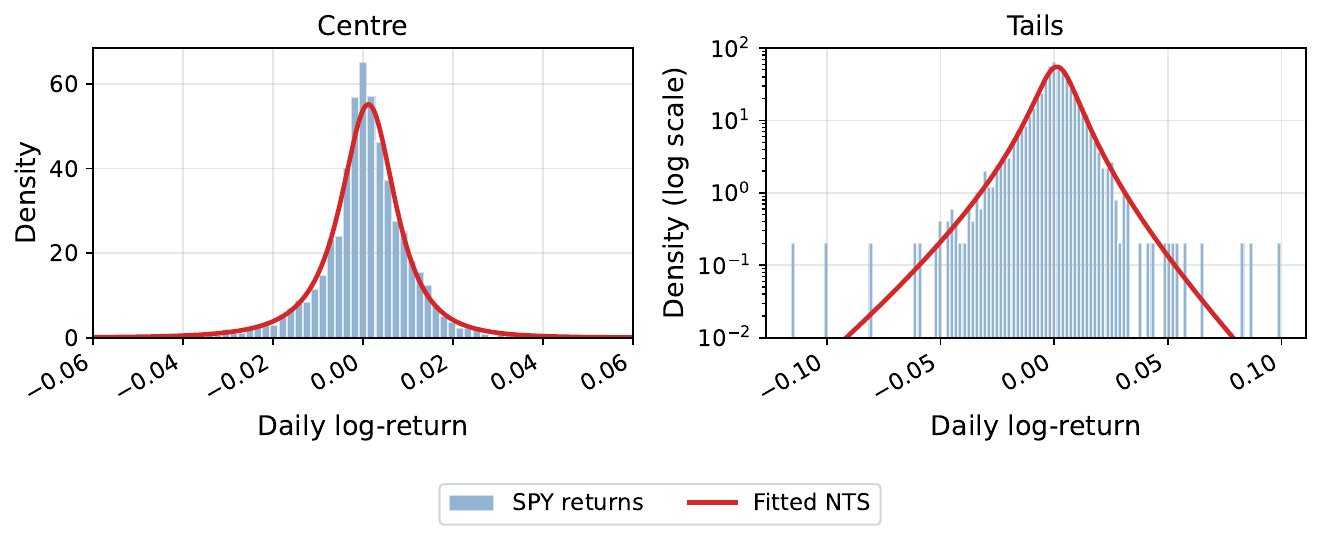}
\caption{SPY daily log-returns and the fitted NTS density: centre of the distribution on a linear scale (left) and tails on a logarithmic scale (right).}
\label{fig:nts_fit}
\end{figure}

Fig.~\ref{fig:nts_fit} compares the fitted density with the histogram of returns on a linear scale in the centre and on a logarithmic scale in the tails. The fit tracks the peak and the body closely and follows the tails to about $\pm 5\%$, or four standard deviations, beyond which the histogram is too sparse for comparison. Formal tests tell the same story in a less forgiving way: the Kolmogorov--Smirnov statistic is $0.031$ for the NTS fit against $0.107$ for a Gaussian with the same mean and variance, and the Cram\'er--von Mises statistic is $0.51$ against $10.4$. With $N = 2{,}746$ both NTS statistics are nevertheless significant at the 5\% level ($p = 0.011$ and $0.038$), whereas the Gaussian is rejected at any level. The tests treat the returns as independent, which daily equity returns are not; volatility clustering inflates the test statistics, and the rejection of the NTS fit should be read as evidence that an unconditional model leaves some structure unexplained rather than as evidence against the family.

\subsection{Channel-Isolated PWF Analysis}\label{subsec:channel_analysis}

With the fitted distribution as benchmark, we apply Definition~\ref{def:channels} with the same settings as in Section~\ref{sec:cases}: scale factors 1.4 and 0.7, a skewness shift of $\pm 0.5$ and a tail shift of $\pm 0.3$. The scale channel thus has $\gamma = 0.01566$ (fearful) and $0.00783$ (greedy). The skew channel moves the skewness from $-0.53$ to $-1.03$ (fearful, $\beta = -0.00170$, $\gamma = 0.01097$) and to $-0.03$ (greedy, $\beta = -0.00005$, $\gamma = 0.01126$), with the standard deviation held at $0.01126$. The tail channel sets $\alpha = 0.77$ (fearful, excess kurtosis 9.6) and $\alpha = 1.37$ (greedy, excess kurtosis 4.9), with $\gamma$ adjusted by less than $0.3\%$ to hold the variance.

\begin{figure}[!tbp]
\centering
\includegraphics[width=0.7\textwidth]{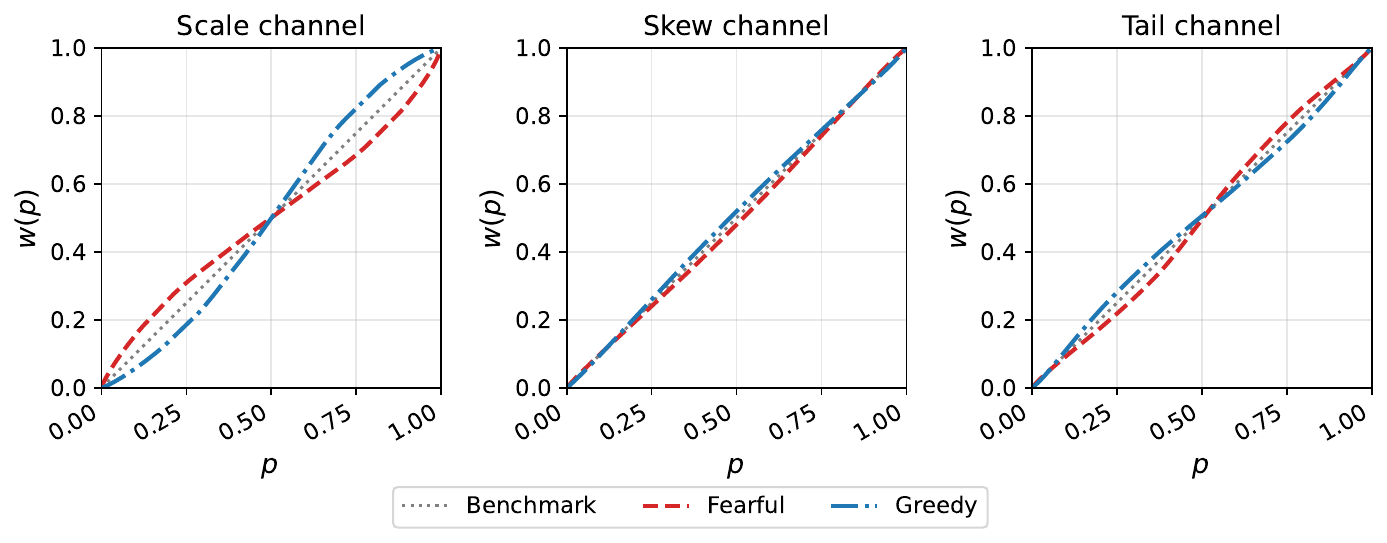}
\caption{Probability-weighting functions for the three channels around the NTS distribution fitted to SPY returns: scale (left), skew (centre), tail (right).}
\label{fig:pwf_channels}
\end{figure}

\begin{figure}[!tbp]
\centering
\includegraphics[width=0.7\textwidth]{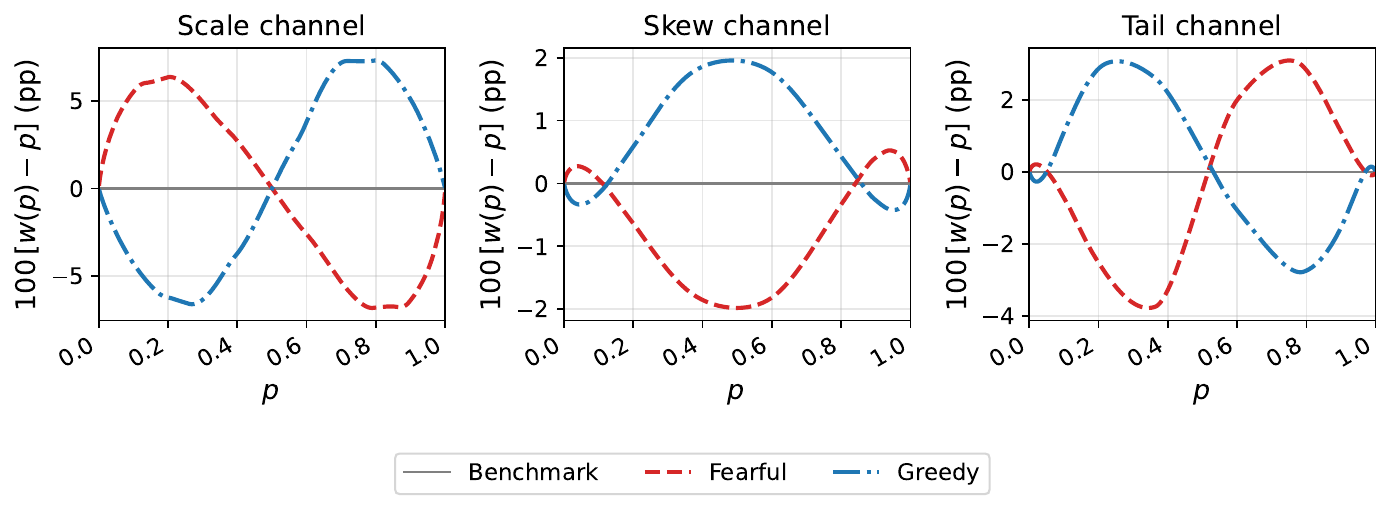}
\caption{PWF deviations $\delta(p) = w(p) - p$ in percentage points for the three channels around the fitted distribution.}
\label{fig:pwf_deviations}
\end{figure}

Fig.~\ref{fig:pwf_channels} plots the PWFs and Fig.~\ref{fig:pwf_deviations} the deviations in percentage points. The scale channel (left panels) gives the inverse-S and S shapes of Case 1, with largest deviations of $6.8$ and $7.3$ percentage points; the crossing point is slightly above $p = 1/2$ because the benchmark is skewed. The skew channel (centre panels) gives one-signed deviations with a largest magnitude of $2.0$ percentage points at the median and small reversals near the endpoints, as in Case 2. The tail channel (right panels) gives deviations with three sign changes and largest magnitudes of $3.8$ (fearful) and $3.1$ (greedy) percentage points, larger than in Case 3 because the fitted benchmark has far heavier tails than the stylized one, so that a change of $0.3$ in $\alpha$ moves the excess kurtosis by about $2.3$ units in either direction rather than by about one.

\subsection{Indices from the Empirical Calibration}\label{subsec:empirical_indices}

\begin{figure}[!tbp]
\centering
\includegraphics[width=0.7\textwidth]{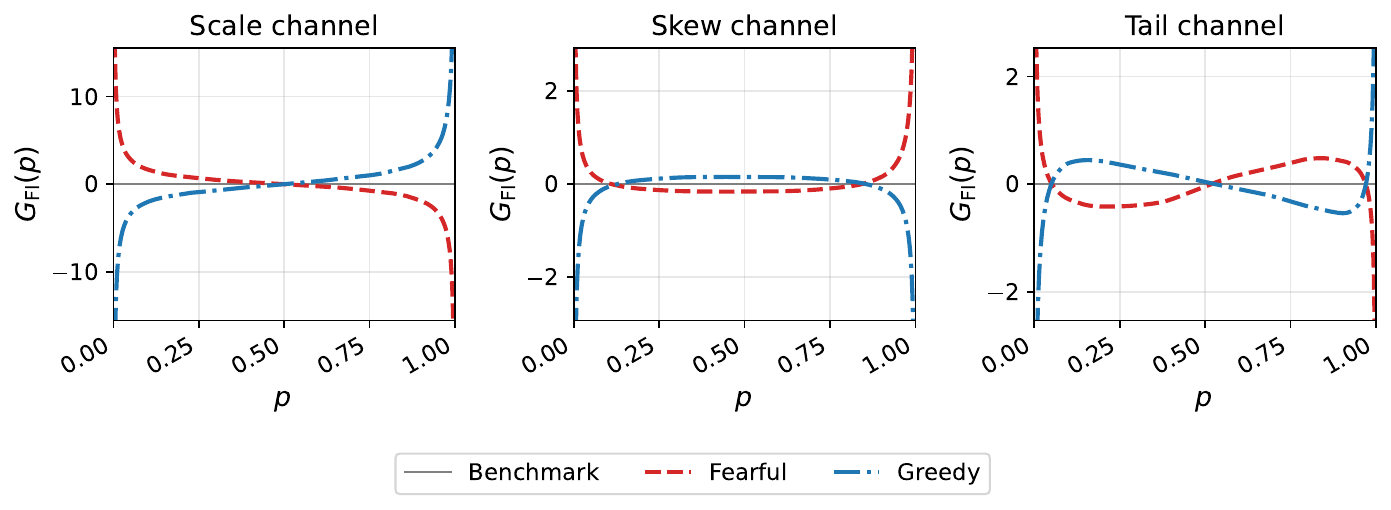}
\caption{Fisher-normalized logit-shift index $G_{\text{FI}}(p)$ for each channel around the fitted distribution: scale (left), skew (centre), tail (right).}
\label{fig:gfi_channels}
\end{figure}

Fig.~\ref{fig:gfi_channels} plots $G_{\text{FI}}(p)$ for each channel, with the axes cut at the range attained on $[0.01, 0.99]$. In the scale channel the index reaches $8.7$ (fearful) and $-10.7$ (greedy) at $p = 0.01$ and is near zero in the centre; in the skew channel it is below $0.3$ in absolute value over the central 80\% of the range and rises to about $2$ at the endpoints; in the tail channel it stays below $0.5$ in the interior and crosses zero three times.

\begin{figure}[!tbp]
\centering
\includegraphics[width=0.7\textwidth]{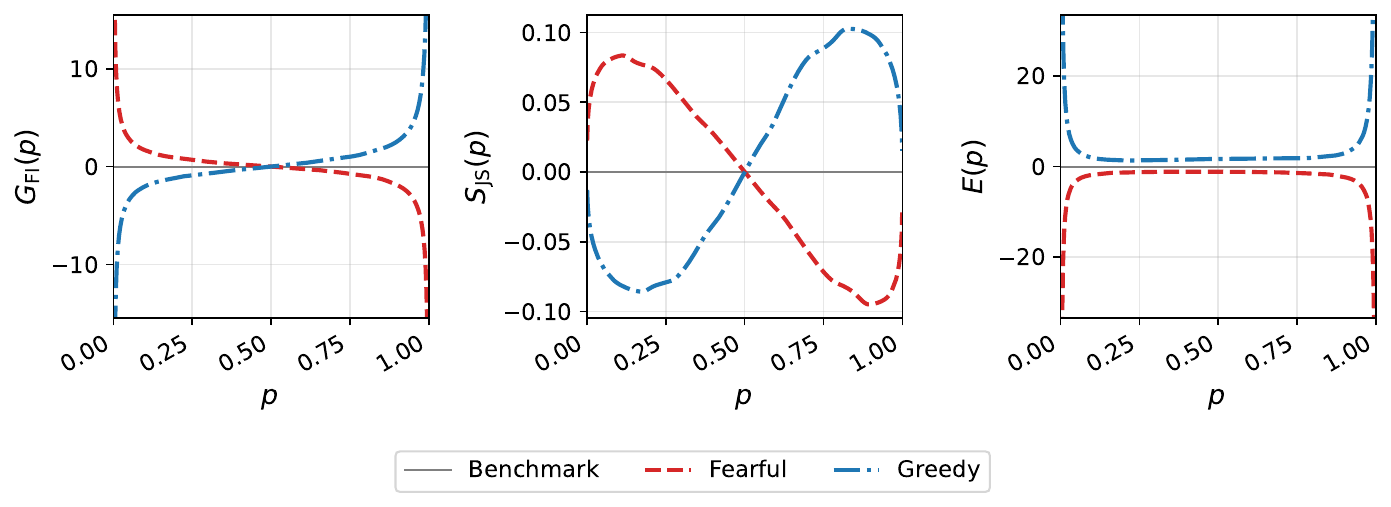}
\caption{The three indices for the scale channel around the fitted distribution: $G_{\text{FI}}(p)$ (left), $S_{\text{JS}}(p)$ (centre), $E(p)$ (right).}
\label{fig:all_indices}
\end{figure}

Fig.~\ref{fig:all_indices} compares the three indices for the scale channel. The logit shift amplifies the tails, the signed Jensen--Shannon index stays within $\pm 0.10$ and peaks at $p \approx 0.2$ and $0.8$, where the deviation itself peaks, and the elasticity is one-signed over the interior (negative for the fearful specification, positive for the greedy one, with $|E| \leq 3$ on $[0.05, 0.95]$) and diverges at the endpoints. The three indices thus read the same PWF at three resolutions: $G_{\text{FI}}$ for the tails, $S_{\text{JS}}$ for the overall size and direction, and $E$ for the local rate of distortion.

Table~\ref{tab:aggregate_indices} reports the largest deviation and the integrated indices \eqref{eq:aggregates} for the six empirical specifications. By every absolute measure the scale channel produces the largest distortions, the tail channel the next largest and the skew channel the smallest; the signed aggregates are small for the scale and tail channels, whose deviations are nearly antisymmetric, and take opposite signs for $\mathcal{G}$ and $\mathcal{S}$ in the skew channel for the reason given in Case 2. The ordering and the sign patterns match those of Section~\ref{sec:cases}, so the stylized results are not an artefact of the benchmark chosen there.

\begin{table}[htbp]
\centering
\caption{Largest deviation and integrated indices for the channels around the fitted distribution}
\label{tab:aggregate_indices}
\begin{tabular}{llcccccc}
\toprule
Channel & Specification & $\max_p|\delta(p)|$ (pp) & $\mathcal{G}$ & $|\mathcal{G}|_1$ & $\mathcal{S}$ & $|\mathcal{S}|_1$ & $\mathcal{E}$ \\
\midrule
Scale & Fearful & 6.8 & $-0.10$ & 1.48 & $-0.003$ & 0.057 & 3.15 \\
Scale & Greedy & 7.3 & $+0.25$ & 1.98 & $+0.005$ & 0.065 & 4.36 \\
Skew & Fearful & 2.0 & $+0.11$ & 0.29 & $-0.008$ & 0.012 & 1.16 \\
Skew & Greedy & 2.0 & $-0.10$ & 0.27 & $+0.008$ & 0.011 & 1.11 \\
Tail & Fearful & 3.8 & $+0.02$ & 0.37 & $-0.001$ & 0.023 & 1.33 \\
Tail & Greedy & 3.1 & $-0.03$ & 0.39 & $+0.002$ & 0.021 & 1.55 \\
\bottomrule
\end{tabular}
\end{table}

A deviation of 7 percentage points is large in decision terms. For the fearful scale specification, an outcome to which the benchmark assigns cumulative probability $0.20$ is weighted as if its cumulative probability were $0.27$; for the greedy specification, an outcome at the $0.80$ quantile is weighted as if it were at $0.87$. These are the magnitudes that a model of option demand or of portfolio choice with probability weighting would have to accommodate if beliefs about volatility moved by 40\%, and they are several times larger than what equally sized changes in skewness or tail thickness produce.

\FloatBarrier

\section{Discussion and Conclusion}\label{sec:discussion}

This paper has constructed probability-weighting functions from changes in the distribution of returns within the normal tempered-stable family and has decomposed the resulting distortions into three channels, scale, skewness and tail thickness, each isolated at constant variance. The decomposition rests on the closed-form moments \eqref{eq:var}--\eqref{eq:kurt}, which make exact variance matching possible, and on the invariance of the quantile mapping to location and scale, which reduces every PWF to a comparison of shapes.

The main computational finding is that the three channels have different signatures and different sizes. Around a symmetric benchmark with excess kurtosis 3, a 40\% change in scale produces an S- or inverse-S-shaped PWF with one crossing of the diagonal and deviations of up to 7 to 8 percentage points; a change in skewness of $\pm 0.5$ produces a one-signed deviation of 2 percentage points that peaks at the median and reverses only in the far tails, giving two crossings; a change of $\pm 0.3$ in the tail index produces a deviation of about 2 percentage points with three crossings. The same patterns, with magnitudes of 7, 2 and 3 to 4 percentage points, appear around parameters fitted to eleven years of S\&P 500 returns. The number of sign changes of $w(p) - p$ is therefore a diagnostic for which feature of the return distribution a given weighting function is responding to, and the magnitudes say that, for perturbations of comparable economic size, beliefs about dispersion matter several times more for probability weights than beliefs about asymmetry or tails.

The three indices complement one another and the classical ratio. The classical index $g(p)$ and the Fisher-normalized logit shift $G_{\text{FI}}(p)$ both diverge at the endpoints whenever the tails of prior and posterior differ, but $G_{\text{FI}}$ diverges more slowly and is measured in units of a single observation's standard error, which makes its values comparable across $p$. The signed Jensen--Shannon index is bounded by $\sqrt{2\log 2}$, vanishes at the endpoints and peaks where the deviation peaks, so it is the natural index for comparing the overall size and direction of distortions across channels, assets or dates. The elasticity locates the inflection points of the log-odds gap. Integrated versions of the three indices give scalar summaries; the signed integrals vanish for symmetric channels and should be read together with their absolute counterparts.

The channel decomposition also has a reading under the risk-neutral measure. The same three parameters that generate behavioral distortions of the physical distribution are the ones that risk-neutral calibrations of NTS models move to fit option prices: scale sets the level of implied volatility, asymmetry the risk reversal, and tail thickness the price of far out-of-the-money options. Comparing the channels estimated under the two measures would decompose the difference between risk-neutral and physical distributions into the same three components, with the indices of Section~\ref{subsec:info-indices} as the metric. We do not carry out that comparison here.

Several limitations qualify the results. The empirical benchmark is estimated by a minimum-distance criterion in the distribution-function domain rather than by maximum likelihood, and the fitted model, while far closer to the data than a Gaussian, is rejected at the 5\% level by tests that assume independent observations. The channel sizes, a 40\% change in scale, a skewness shift of $0.5$ and a tail-index shift of $0.3$, are choices; the signatures do not depend on them, but the magnitudes do, roughly in proportion. The analysis is unconditional and static: it compares two distributions, not a process in which beliefs move over time. And the economic consequences of the distortions are stated in terms of probability weights only; translating them into prices or portfolio weights requires a model of preferences and of the market that is outside the scope of this paper.

These limitations point to extensions. Estimating the NTS parameters by maximum likelihood or by a characteristic-function method, and allowing for volatility clustering through a GARCH or stochastic-volatility filter before fitting the unconditional law, would give a benchmark on which the goodness-of-fit tests are better founded. Calibrating the same model to option prices and comparing the physical and risk-neutral parameters channel by channel would connect the indices to risk premia. Rolling estimation of $S_{\text{JS}}$ and of the integrated indices would turn them into sentiment series whose forecasting content could be tested. Multivariate NTS laws with factor or copula dependence would extend the construction to portfolios. And choice experiments in which the return distribution is varied along one channel at a time would test the signatures directly.

In summary, the quantile mapping $w(p) = F_{\text{post}}(Q_{\text{prior}}(p))$ turns a change in beliefs about returns into a probability-weighting function with a known cause; within the NTS family the causes can be separated into scale, skewness and tail thickness at constant variance; each channel leaves a distinct signature in the number of sign changes of $w(p) - p$; and the information-theoretic indices introduced here measure the resulting distortions without the endpoint singularity of the classical ratio. All figures and numbers in the paper are reproducible from the accompanying code.

\backmatter


\section*{Declarations}

\begin{description}
\item[Funding:] Not applicable.
\item[Conflict of Interest:] The authors declare no conflicts of interest.
\item[Ethics Approval:] Not applicable.
\item[Consent to Participate:] Not applicable.
\item[Consent for Publication:] Not applicable.
\item[Data Availability:] SPY price data were obtained from Bloomberg. The daily log-return series used in Section~\ref{sec:empirical} is included with the code.
\item[Code Availability:] Python code that reproduces every figure, table and number in the paper is available at \url{https://github.com/akashdeepo/NTS_PWF}.
\item[Use of Generative AI:] The authors used Claude (Anthropic) to assist with debugging and verifying the code, revising the manuscript for clarity and readability, and preparing the public code repository. The authors reviewed all output, checked every reported result, and take full responsibility for the content of the article.
\item[Author Contributions:] Conceptualization: STR; Methodology: AD, BD, STR; Software: AD, BD, AT; Formal Analysis: AD, BD; Investigation: AD, BD; Resources: STR, WBL, FF; Writing - Original Draft: AD; Writing - Review \& Editing: WBL, FF, STR; Visualization: AD, BD; Supervision: STR, WBL, FF; Project Administration: STR. All authors have read and agreed to the published version of the manuscript.
\end{description}

\appendix

\section{Numerical Details}\label{app:numerics}

\subsection{Fourier Inversion}\label{app:inversion}

All computations use the \texttt{temStaPy} implementation of the NTS distribution. Each distribution is first standardized to zero mean and unit variance, since the characteristic function of a standardized NTS law depends only on $(\alpha, \theta, \beta/\sigma)$, and the location and scale are restored afterwards. The density is computed by applying the fast Fourier transform to $\phi_X$ sampled at $2^{17}$ points with spacing $2^{-10}$, and the distribution function by applying it to the integrand of \eqref{eq:gilpelaez} sampled at $2^{12}$ points with spacing $2^{-8}$. Values between grid points are obtained by monotone piecewise-cubic interpolation, which preserves monotonicity of the distribution function. Negative density values produced by truncation error in the far tails are set to zero.

\subsection{Quantile Function and PWF}\label{app:quantile}

The quantile function is obtained by tabulating the standardized distribution function on $[-10, 10]$ with step $0.01$, removing any non-increasing values, and interpolating the inverse by the same monotone cubic scheme. The PWF \eqref{eq:pwf} is then $F_{\text{post}}$ evaluated at the prior quantiles of the grid $p_i = i/2001$, $i = 1, \dots, 2000$, and its derivative is the density ratio \eqref{eq:wprime} at the same points. Probabilities entering logarithms are clipped to $[10^{-6}, 1 - 10^{-6}]$. The integrated indices \eqref{eq:aggregates} are computed by the trapezoidal rule on the same grid.

\subsection{Channel Construction}\label{app:channels}

For the skew channel, $\beta$ is found by bracketing and bisection (Brent's method) on the skewness formula \eqref{eq:skew} with $\gamma$ eliminated through the variance constraint \eqref{eq:var}; the admissible range is $|\beta| < \sigma/\sqrt{v}$. For the tail channel, $\gamma$ is reset to $(\sigma^2 - \beta^2 v)^{1/2}$ with the new value of $v$. The closed-form moments \eqref{eq:mean}--\eqref{eq:kurt} agree with the cumulant routine of \texttt{temStaPy} to machine precision, and the characteristic function \eqref{eq:cf} agrees with the library's to the same accuracy.


\bibliography{nts_pwf_references}

\begin{thebibliography}{}
\renewcommand{\doi}[1]{\url{https://doi.org/#1}}
\bibcommenthead

\bibitem [\protect \citeauthoryear {%
Applebaum%
}{%
Applebaum%
}{%
{\protect \APACyear {2009}}%
}]{%
applebaum2009levy}
\APACinsertmetastar {%
applebaum2009levy}%
\begin{APACrefauthors}%
Applebaum, D.%
\end{APACrefauthors}%
\unskip\
\newblock
\APACrefYear{2009}.
\newblock
\APACrefbtitle {{L}\'evy Processes and Stochastic Calculus} {{L}\'evy processes
  and stochastic calculus}\ (\PrintOrdinal{2nd}\ \BEd).
\newblock
\APACaddressPublisher{Cambridge}{Cambridge University Press}.
\PrintBackRefs{\CurrentBib}

\bibitem [\protect \citeauthoryear {%
Bakshi%
, Carr%
\BCBL {}\ \protect \BOthers {.}}{%
Bakshi%
, Carr%
\BCBL {}\ \protect \BOthers {.}}{%
{\protect \APACyear {2008}}%
}]{%
bakshi2008stochastic}
\APACinsertmetastar {%
bakshi2008stochastic}%
\begin{APACrefauthors}%
Bakshi, G.%
, Carr, P.%
\BCBL {} Wu, L.%
\end{APACrefauthors}%
\unskip\
\newblock
\APACrefYearMonthDay{2008}{}{}.
\newblock
{\BBOQ}\APACrefatitle {Stochastic risk premiums, stochastic skewness in
  currency options, and stochastic discount factors in international economies}
  {Stochastic risk premiums, stochastic skewness in currency options, and
  stochastic discount factors in international economies}.{\BBCQ}
\newblock
\APACjournalVolNumPages{Journal of Financial Economics}{87}{1}{132--156,}
\newblock
\begin{APACrefDOI} \doi{10.1016/j.jfineco.2006.12.001} \end{APACrefDOI}
\newblock

\newblock

\PrintBackRefs{\CurrentBib}

\bibitem [\protect \citeauthoryear {%
Barberis%
\ \BBA {} Huang%
}{%
Barberis%
\ \BBA {} Huang%
}{%
{\protect \APACyear {2008}}%
}]{%
barberis2008stocks}
\APACinsertmetastar {%
barberis2008stocks}%
\begin{APACrefauthors}%
Barberis, N.%
\BCBT {}\ \BBA {} Huang, M.%
\end{APACrefauthors}%
\unskip\
\newblock
\APACrefYearMonthDay{2008}{}{}.
\newblock
{\BBOQ}\APACrefatitle {Stocks as lotteries: The implications of probability
  weighting for security prices} {Stocks as lotteries: The implications of
  probability weighting for security prices}.{\BBCQ}
\newblock
\APACjournalVolNumPages{American Economic Review}{98}{5}{2066--2100,}
\newblock
\begin{APACrefDOI} \doi{10.1257/aer.98.5.2066} \end{APACrefDOI}
\newblock

\newblock

\PrintBackRefs{\CurrentBib}

\bibitem [\protect \citeauthoryear {%
Barndorff-Nielsen%
}{%
Barndorff-Nielsen%
}{%
{\protect \APACyear {1997a}}%
}]{%
barndorff1997normal}
\APACinsertmetastar {%
barndorff1997normal}%
\begin{APACrefauthors}%
Barndorff-Nielsen, O.E.%
\end{APACrefauthors}%
\unskip\
\newblock
\APACrefYearMonthDay{1997a}{}{}.
\newblock
{\BBOQ}\APACrefatitle {Normal inverse {G}aussian distributions and stochastic
  volatility modelling} {Normal inverse {G}aussian distributions and stochastic
  volatility modelling}.{\BBCQ}
\newblock
\APACjournalVolNumPages{Scandinavian Journal of Statistics}{24}{1}{1--13,}
\newblock
\begin{APACrefDOI} \doi{10.1111/1467-9469.00045} \end{APACrefDOI}
\newblock

\newblock

\PrintBackRefs{\CurrentBib}

\bibitem [\protect \citeauthoryear {%
Barndorff-Nielsen%
}{%
Barndorff-Nielsen%
}{%
{\protect \APACyear {1997b}}%
}]{%
barndorff1997processes}
\APACinsertmetastar {%
barndorff1997processes}%
\begin{APACrefauthors}%
Barndorff-Nielsen, O.E.%
\end{APACrefauthors}%
\unskip\
\newblock
\APACrefYearMonthDay{1997b}{}{}.
\newblock
{\BBOQ}\APACrefatitle {Processes of normal inverse {G}aussian type} {Processes
  of normal inverse {G}aussian type}.{\BBCQ}
\newblock
\APACjournalVolNumPages{Finance and Stochastics}{2}{1}{41--68,}
\newblock
\begin{APACrefDOI} \doi{10.1007/s007800050032} \end{APACrefDOI}
\newblock

\newblock

\PrintBackRefs{\CurrentBib}

\bibitem [\protect \citeauthoryear {%
Bochner%
}{%
Bochner%
}{%
{\protect \APACyear {1949}}%
}]{%
bochner1949diffusion}
\APACinsertmetastar {%
bochner1949diffusion}%
\begin{APACrefauthors}%
Bochner, S.%
\end{APACrefauthors}%
\unskip\
\newblock
\APACrefYearMonthDay{1949}{}{}.
\newblock
{\BBOQ}\APACrefatitle {Diffusion equation and stochastic processes} {Diffusion
  equation and stochastic processes}.{\BBCQ}
\newblock
\APACjournalVolNumPages{Proceedings of the National Academy of
  Sciences}{35}{7}{368--370,}
\newblock
\begin{APACrefDOI} \doi{10.1073/pnas.35.7.368} \end{APACrefDOI}
\newblock

\newblock

\PrintBackRefs{\CurrentBib}

\bibitem [\protect \citeauthoryear {%
Carr%
, Geman%
\BCBL {}\ \protect \BOthers {.}}{%
Carr%
, Geman%
\BCBL {}\ \protect \BOthers {.}}{%
{\protect \APACyear {2002}}%
}]{%
carr2002fine}
\APACinsertmetastar {%
carr2002fine}%
\begin{APACrefauthors}%
Carr, P.%
, Geman, H.%
, Madan, D.B.%
\BCBL {} Yor, M.%
\end{APACrefauthors}%
\unskip\
\newblock
\APACrefYearMonthDay{2002}{}{}.
\newblock
{\BBOQ}\APACrefatitle {The fine structure of asset returns: An empirical
  investigation} {The fine structure of asset returns: An empirical
  investigation}.{\BBCQ}
\newblock
\APACjournalVolNumPages{The Journal of Business}{75}{2}{305--332,}
\newblock
\begin{APACrefDOI} \doi{10.1086/338705} \end{APACrefDOI}
\newblock

\newblock

\PrintBackRefs{\CurrentBib}

\bibitem [\protect \citeauthoryear {%
Carr%
\ \BBA {} Madan%
}{%
Carr%
\ \BBA {} Madan%
}{%
{\protect \APACyear {1999}}%
}]{%
carr1999option}
\APACinsertmetastar {%
carr1999option}%
\begin{APACrefauthors}%
Carr, P.%
\BCBT {}\ \BBA {} Madan, D.B.%
\end{APACrefauthors}%
\unskip\
\newblock
\APACrefYearMonthDay{1999}{}{}.
\newblock
{\BBOQ}\APACrefatitle {Option valuation using the fast {F}ourier transform}
  {Option valuation using the fast {F}ourier transform}.{\BBCQ}
\newblock
\APACjournalVolNumPages{Journal of Computational Finance}{2}{4}{61--73,}
\newblock
\begin{APACrefDOI} \doi{10.21314/jcf.1999.043} \end{APACrefDOI}
\newblock

\newblock

\PrintBackRefs{\CurrentBib}

\bibitem [\protect \citeauthoryear {%
Cont%
}{%
Cont%
}{%
{\protect \APACyear {2001}}%
}]{%
cont2001empirical}
\APACinsertmetastar {%
cont2001empirical}%
\begin{APACrefauthors}%
Cont, R.%
\end{APACrefauthors}%
\unskip\
\newblock
\APACrefYearMonthDay{2001}{}{}.
\newblock
{\BBOQ}\APACrefatitle {Empirical properties of asset returns: {S}tylized facts
  and statistical issues} {Empirical properties of asset returns: {S}tylized
  facts and statistical issues}.{\BBCQ}
\newblock
\APACjournalVolNumPages{Quantitative Finance}{1}{2}{223--236,}
\newblock
\begin{APACrefDOI} \doi{10.1080/713665670} \end{APACrefDOI}
\newblock

\newblock

\PrintBackRefs{\CurrentBib}

\bibitem [\protect \citeauthoryear {%
Cont%
\ \BBA {} Tankov%
}{%
Cont%
\ \BBA {} Tankov%
}{%
{\protect \APACyear {2003}}%
}]{%
cont2003financial}
\APACinsertmetastar {%
cont2003financial}%
\begin{APACrefauthors}%
Cont, R.%
\BCBT {}\ \BBA {} Tankov, P.%
\end{APACrefauthors}%
\unskip\
\newblock
\APACrefYear{2003}.
\newblock
\APACrefbtitle {Financial Modelling with Jump Processes} {Financial modelling
  with jump processes}.
\newblock
\APACaddressPublisher{Boca Raton}{Chapman and Hall/CRC}.
\PrintBackRefs{\CurrentBib}

\bibitem [\protect \citeauthoryear {%
Csisz{\'a}r%
}{%
Csisz{\'a}r%
}{%
{\protect \APACyear {1967}}%
}]{%
csiszar1967information}
\APACinsertmetastar {%
csiszar1967information}%
\begin{APACrefauthors}%
Csisz{\'a}r, I.%
\end{APACrefauthors}%
\unskip\
\newblock
\APACrefYearMonthDay{1967}{}{}.
\newblock
{\BBOQ}\APACrefatitle {Information-type measures of difference of probability
  distributions and indirect observations} {Information-type measures of
  difference of probability distributions and indirect observations}.{\BBCQ}
\newblock
\APACjournalVolNumPages{Studia Scientiarum Mathematicarum
  Hungarica}{2}{}{299--318,}
\newblock

\newblock

\PrintBackRefs{\CurrentBib}

\bibitem [\protect \citeauthoryear {%
Deep%
, Rachev%
\BCBL {}\ \protect \BOthers {.}}{%
Deep%
, Rachev%
\BCBL {}\ \protect \BOthers {.}}{%
{\protect \APACyear {2025}}%
}]{%
deep2025probability}
\APACinsertmetastar {%
deep2025probability}%
\begin{APACrefauthors}%
Deep, A.%
, Rachev, S.T.%
\BCBL {} Fabozzi, F.J.%
\end{APACrefauthors}%
\unskip\
\newblock
\APACrefYearMonthDay{2025}{}{}.
\newblock
\APACrefbtitle {Probability Weighting Meets Heavy Tails: An Econometric
  Framework for Behavioral Asset Pricing.} {Probability weighting meets heavy
  tails: An econometric framework for behavioral asset pricing.}
\newblock
\APAChowpublished {arXiv preprint arXiv:2511.16563}.
\PrintBackRefs{\CurrentBib}

\bibitem [\protect \citeauthoryear {%
Endres%
\ \BBA {} Schindelin%
}{%
Endres%
\ \BBA {} Schindelin%
}{%
{\protect \APACyear {2003}}%
}]{%
endres2003new}
\APACinsertmetastar {%
endres2003new}%
\begin{APACrefauthors}%
Endres, D.M.%
\BCBT {}\ \BBA {} Schindelin, J.E.%
\end{APACrefauthors}%
\unskip\
\newblock
\APACrefYearMonthDay{2003}{}{}.
\newblock
{\BBOQ}\APACrefatitle {A new metric for probability distributions} {A new
  metric for probability distributions}.{\BBCQ}
\newblock
\APACjournalVolNumPages{IEEE Transactions on Information
  Theory}{49}{7}{1858--1860,}
\newblock
\begin{APACrefDOI} \doi{10.1109/tit.2003.813506} \end{APACrefDOI}
\newblock

\newblock

\PrintBackRefs{\CurrentBib}

\bibitem [\protect \citeauthoryear {%
Gil-Pelaez%
}{%
Gil-Pelaez%
}{%
{\protect \APACyear {1951}}%
}]{%
gilpelaez1951note}
\APACinsertmetastar {%
gilpelaez1951note}%
\begin{APACrefauthors}%
Gil-Pelaez, J.%
\end{APACrefauthors}%
\unskip\
\newblock
\APACrefYearMonthDay{1951}{}{}.
\newblock
{\BBOQ}\APACrefatitle {Note on the inversion theorem} {Note on the inversion
  theorem}.{\BBCQ}
\newblock
\APACjournalVolNumPages{Biometrika}{38}{3--4}{481--482,}
\newblock
\begin{APACrefDOI} \doi{10.2307/2332598} \end{APACrefDOI}
\newblock

\newblock

\PrintBackRefs{\CurrentBib}

\bibitem [\protect \citeauthoryear {%
Harvey%
\ \BBA {} Siddique%
}{%
Harvey%
\ \BBA {} Siddique%
}{%
{\protect \APACyear {1999}}%
}]{%
harvey1999autoregressive}
\APACinsertmetastar {%
harvey1999autoregressive}%
\begin{APACrefauthors}%
Harvey, C.R.%
\BCBT {}\ \BBA {} Siddique, A.%
\end{APACrefauthors}%
\unskip\
\newblock
\APACrefYearMonthDay{1999}{}{}.
\newblock
{\BBOQ}\APACrefatitle {Autoregressive conditional skewness} {Autoregressive
  conditional skewness}.{\BBCQ}
\newblock
\APACjournalVolNumPages{Journal of Financial and Quantitative
  Analysis}{34}{4}{465--487,}
\newblock
\begin{APACrefDOI} \doi{10.2307/2676230} \end{APACrefDOI}
\newblock

\newblock

\PrintBackRefs{\CurrentBib}

\bibitem [\protect \citeauthoryear {%
Kahneman%
\ \BBA {} Tversky%
}{%
Kahneman%
\ \BBA {} Tversky%
}{%
{\protect \APACyear {1979}}%
}]{%
kahneman1979prospect}
\APACinsertmetastar {%
kahneman1979prospect}%
\begin{APACrefauthors}%
Kahneman, D.%
\BCBT {}\ \BBA {} Tversky, A.%
\end{APACrefauthors}%
\unskip\
\newblock
\APACrefYearMonthDay{1979}{}{}.
\newblock
{\BBOQ}\APACrefatitle {Prospect theory: An analysis of decision under risk}
  {Prospect theory: An analysis of decision under risk}.{\BBCQ}
\newblock
\APACjournalVolNumPages{Econometrica}{47}{2}{263--291,}
\newblock
\begin{APACrefDOI} \doi{10.2307/1914185} \end{APACrefDOI}
\newblock

\newblock

\PrintBackRefs{\CurrentBib}

\bibitem [\protect \citeauthoryear {%
Kim%
, Rachev%
\BCBL {}\ \protect \BOthers {.}}{%
Kim%
, Rachev%
\BCBL {}\ \protect \BOthers {.}}{%
{\protect \APACyear {2010}}%
}]{%
kim2010tempered}
\APACinsertmetastar {%
kim2010tempered}%
\begin{APACrefauthors}%
Kim, Y.S.%
, Rachev, S.T.%
, Bianchi, M.L.%
\BCBL {} Fabozzi, F.J.%
\end{APACrefauthors}%
\unskip\
\newblock
\APACrefYearMonthDay{2010}{}{}.
\newblock
{\BBOQ}\APACrefatitle {Tempered stable and tempered infinitely divisible
  {GARCH} models} {Tempered stable and tempered infinitely divisible {GARCH}
  models}.{\BBCQ}
\newblock
\APACjournalVolNumPages{Journal of Banking \& Finance}{34}{9}{2096--2109,}
\newblock
\begin{APACrefDOI} \doi{10.1016/j.jbankfin.2010.01.015} \end{APACrefDOI}
\newblock

\newblock

\PrintBackRefs{\CurrentBib}

\bibitem [\protect \citeauthoryear {%
Lin%
}{%
Lin%
}{%
{\protect \APACyear {1991}}%
}]{%
lin1991divergence}
\APACinsertmetastar {%
lin1991divergence}%
\begin{APACrefauthors}%
Lin, J.%
\end{APACrefauthors}%
\unskip\
\newblock
\APACrefYearMonthDay{1991}{}{}.
\newblock
{\BBOQ}\APACrefatitle {Divergence measures based on the {S}hannon entropy}
  {Divergence measures based on the {S}hannon entropy}.{\BBCQ}
\newblock
\APACjournalVolNumPages{IEEE Transactions on Information
  Theory}{37}{1}{145--151,}
\newblock
\begin{APACrefDOI} \doi{10.1109/18.61115} \end{APACrefDOI}
\newblock

\newblock

\PrintBackRefs{\CurrentBib}

\bibitem [\protect \citeauthoryear {%
Madan%
, Carr%
\BCBL {}\ \protect \BOthers {.}}{%
Madan%
, Carr%
\BCBL {}\ \protect \BOthers {.}}{%
{\protect \APACyear {1998}}%
}]{%
madan1998variance}
\APACinsertmetastar {%
madan1998variance}%
\begin{APACrefauthors}%
Madan, D.B.%
, Carr, P.P.%
\BCBL {} Chang, E.C.%
\end{APACrefauthors}%
\unskip\
\newblock
\APACrefYearMonthDay{1998}{}{}.
\newblock
{\BBOQ}\APACrefatitle {The variance gamma process and option pricing} {The
  variance gamma process and option pricing}.{\BBCQ}
\newblock
\APACjournalVolNumPages{European Finance Review}{2}{1}{79--105,}
\newblock
\begin{APACrefDOI} \doi{10.1023/a:1009703431535} \end{APACrefDOI}
\newblock

\newblock

\PrintBackRefs{\CurrentBib}

\bibitem [\protect \citeauthoryear {%
Madan%
\ \BBA {} Yor%
}{%
Madan%
\ \BBA {} Yor%
}{%
{\protect \APACyear {2008}}%
}]{%
madan2008representing}
\APACinsertmetastar {%
madan2008representing}%
\begin{APACrefauthors}%
Madan, D.B.%
\BCBT {}\ \BBA {} Yor, M.%
\end{APACrefauthors}%
\unskip\
\newblock
\APACrefYearMonthDay{2008}{}{}.
\newblock
{\BBOQ}\APACrefatitle {Representing the {CGMY} and {M}eixner {L}\'evy processes
  as time changed {B}rownian motions} {Representing the {CGMY} and {M}eixner
  {L}\'evy processes as time changed {B}rownian motions}.{\BBCQ}
\newblock
\APACjournalVolNumPages{Journal of Computational Finance}{12}{1}{27--47,}
\newblock
\begin{APACrefDOI} \doi{10.21314/jcf.2008.181} \end{APACrefDOI}
\newblock

\newblock

\PrintBackRefs{\CurrentBib}

\bibitem [\protect \citeauthoryear {%
Phillips%
}{%
Phillips%
}{%
{\protect \APACyear {1952}}%
}]{%
phillips1952generation}
\APACinsertmetastar {%
phillips1952generation}%
\begin{APACrefauthors}%
Phillips, R.S.%
\end{APACrefauthors}%
\unskip\
\newblock
\APACrefYearMonthDay{1952}{}{}.
\newblock
{\BBOQ}\APACrefatitle {On the generation of semigroups of linear operators} {On
  the generation of semigroups of linear operators}.{\BBCQ}
\newblock
\APACjournalVolNumPages{Pacific Journal of Mathematics}{2}{3}{343--369,}
\newblock
\begin{APACrefDOI} \doi{10.2140/pjm.1952.2.343} \end{APACrefDOI}
\newblock

\newblock

\PrintBackRefs{\CurrentBib}

\bibitem [\protect \citeauthoryear {%
Prelec%
}{%
Prelec%
}{%
{\protect \APACyear {1998}}%
}]{%
prelec1998probability}
\APACinsertmetastar {%
prelec1998probability}%
\begin{APACrefauthors}%
Prelec, D.%
\end{APACrefauthors}%
\unskip\
\newblock
\APACrefYearMonthDay{1998}{}{}.
\newblock
{\BBOQ}\APACrefatitle {The probability weighting function} {The probability
  weighting function}.{\BBCQ}
\newblock
\APACjournalVolNumPages{Econometrica}{66}{3}{497--527,}
\newblock
\begin{APACrefDOI} \doi{10.2307/2998573} \end{APACrefDOI}
\newblock

\newblock

\PrintBackRefs{\CurrentBib}

\bibitem [\protect \citeauthoryear {%
Rachev%
, Kim%
\BCBL {}\ \protect \BOthers {.}}{%
Rachev%
, Kim%
\BCBL {}\ \protect \BOthers {.}}{%
{\protect \APACyear {2011}}%
}]{%
rachev2011financial}
\APACinsertmetastar {%
rachev2011financial}%
\begin{APACrefauthors}%
Rachev, S.T.%
, Kim, Y.S.%
, Bianchi, M.L.%
\BCBL {} Fabozzi, F.J.%
\end{APACrefauthors}%
\unskip\
\newblock
\APACrefYear{2011}.
\newblock
\APACrefbtitle {Financial Models with {L}\'evy Processes and Volatility
  Clustering} {Financial models with {L}\'evy processes and volatility
  clustering}.
\newblock
\APACaddressPublisher{Hoboken, NJ}{John Wiley \& Sons}.
\PrintBackRefs{\CurrentBib}

\bibitem [\protect \citeauthoryear {%
Rosi{\'n}ski%
}{%
Rosi{\'n}ski%
}{%
{\protect \APACyear {2007}}%
}]{%
rosinski2007tempering}
\APACinsertmetastar {%
rosinski2007tempering}%
\begin{APACrefauthors}%
Rosi{\'n}ski, J.%
\end{APACrefauthors}%
\unskip\
\newblock
\APACrefYearMonthDay{2007}{}{}.
\newblock
{\BBOQ}\APACrefatitle {Tempering stable processes} {Tempering stable
  processes}.{\BBCQ}
\newblock
\APACjournalVolNumPages{Stochastic Processes and their
  Applications}{117}{6}{677--707,}
\newblock
\begin{APACrefDOI} \doi{10.1016/j.spa.2006.10.003} \end{APACrefDOI}
\newblock

\newblock

\PrintBackRefs{\CurrentBib}

\bibitem [\protect \citeauthoryear {%
Schilling%
, Song%
\BCBL {}\ \protect \BOthers {.}}{%
Schilling%
, Song%
\BCBL {}\ \protect \BOthers {.}}{%
{\protect \APACyear {2012}}%
}]{%
schilling2012bernstein}
\APACinsertmetastar {%
schilling2012bernstein}%
\begin{APACrefauthors}%
Schilling, R.L.%
, Song, R.%
\BCBL {} Vondra{\v{c}}ek, Z.%
\end{APACrefauthors}%
\unskip\
\newblock
\APACrefYear{2012}.
\newblock
\APACrefbtitle {Bernstein Functions: Theory and Applications} {Bernstein
  functions: Theory and applications}\ (\PrintOrdinal{2nd}\ \BEd, \BVOL~37).
\newblock
\APACaddressPublisher{Berlin}{Walter de Gruyter}.
\PrintBackRefs{\CurrentBib}

\bibitem [\protect \citeauthoryear {%
Tversky%
\ \BBA {} Kahneman%
}{%
Tversky%
\ \BBA {} Kahneman%
}{%
{\protect \APACyear {1992}}%
}]{%
tversky1992advances}
\APACinsertmetastar {%
tversky1992advances}%
\begin{APACrefauthors}%
Tversky, A.%
\BCBT {}\ \BBA {} Kahneman, D.%
\end{APACrefauthors}%
\unskip\
\newblock
\APACrefYearMonthDay{1992}{}{}.
\newblock
{\BBOQ}\APACrefatitle {Advances in prospect theory: Cumulative representation
  of uncertainty} {Advances in prospect theory: Cumulative representation of
  uncertainty}.{\BBCQ}
\newblock
\APACjournalVolNumPages{Journal of Risk and Uncertainty}{5}{4}{297--323,}
\newblock
\begin{APACrefDOI} \doi{10.1007/bf00122574} \end{APACrefDOI}
\newblock

\newblock

\PrintBackRefs{\CurrentBib}

\end{thebibliography}

\end{document}